\documentclass[aps,prl,groupedaddress,showpacs,floatfix,nofootinbib,preprintnumbers,notitlepage,twocolumn]{revtex4-2}
\usepackage[utf8]{inputenc}
\usepackage[T1]{fontenc}
\usepackage{lmodern}
\usepackage{amsmath,amssymb,bm,mathtools,mathrsfs}
\usepackage[letterpaper,top=1.75cm,bottom=1.75cm,left=1.65cm,right=1.65cm,columnsep=0.55cm]{geometry}

\usepackage{graphicx}
\usepackage{float}

\usepackage{color}
\usepackage{multirow}
\usepackage{xcolor}
\usepackage[colorlinks=true,citecolor=blue!55!black,linkcolor=blue!55!black,urlcolor=blue!55!black]{hyperref}
\usepackage{booktabs}
\usepackage{microtype}

\usepackage{orcidlink}
\makeatletter
\def\NAT@def@citea{\def\@citea{\NAT@separator}}
\makeatother

\begin{document}
    
\title{Emergence of the nuclear halo in 
$^{22}$C: Bridging shape and shell}

\author{Xiang-Xiang Sun\orcidlink{0000-0003-2809-4638}}
\email{sunxiangxiang@itp.ac.cn}
\affiliation{Institute of Theoretical Physics, Chinese Academy of Sciences, Beijing 100190, China} 
\affiliation{Institute for Advanced Simulation (IAS-4), Forschungszentrum J\"{u}lich, D-52425 J\"{u}lich, Germany}

\date{\today}
\begin{abstract}
The structure of $^{22}$C remains unsettled more than a decade after the 2010 reaction-cross-section measurement.
The nucleus sits at $N=16$, where the spherical closure of $^{24}$O
competes with the deformation of $^{20}$C.
We develop a multi-reference density functional approach that restores
particle number and angular momentum and mixes quadrupole shapes while
retaining the continuum.
The calculation gives a deformed two-neutron halo with
matter radius of 3.29 fm, closer to the carbon-target extraction,
and a long $s$-wave tail.
The ground state $0_{1}^{+}$ is dominated by oblate configurations, 
with $B(E2)$ values comparable to those of
$^{20}$C.
The valence neutrons occupy mixed $sd$ orbits; the $N=16$ closure
established in $^{24}$O is substantially weakened in $^{22}$C.
\end{abstract}
\maketitle

{\it Introduction.-}~
Nuclear halos, which are weakly bound and spatially extended, remain strongly interacting quantum many-body systems, even when their size and dipole response display an emergent few-body character~\cite{Jensen2004_RMP76-215,Tanihata2013_PPNP68-215,Meng2015_JPG42-093101,Nakamura2019_AAPPSBL29-5}. 
Most nuclei are deformed, and near the drip line weak binding competes with deformation and shell evolution: valence neutrons in low-$\Omega$ orbitals can form a halo whose shape differs from that of the core~\cite{Misu1997_NPA614-44,Hamamoto2010_PRC81-021304,Zhou2010_PRC82-011301R}.
So far, deformed halos have been identified only in one-neutron halo nuclei $^{31}$Ne and $^{37}$Mg \cite{Nakamura2009_PRL103-262501, Nakamura2014_PRL112-142501,Kobayashi2014_PRL112-242501}. 
A small $s$-wave component has also been measured in $^{17}$B and interpreted as a deformed halo \cite{Yang2021_PRL126-082501}.
$^{22}$C \cite{Pougheon1986_EPL2-505}, the heaviest known two-neutron halo \cite{Tanaka2010_PRL104-062701} before the confimation of the halo in $^{29}$F, 
is the prime candidate for a deformed Borromean halo
\cite{Sun2018_PLB785-530}, because its core $^{20}$C is soft enough to change shape \cite{Stanoiu2008_PRC78-034315,Elekes2004_PLB586-34,Petri2011_PRL107-102501}.
Its neutron number 
$N=16$ coincides with the shell closure established in $^{24}$O \cite{Ozawa2000_PRL84-5493,Kanungo2009_PRL102-152501,Tshoo2012_PRL109-22501}, 
so the structure of 
$^{22}$C also tests whether this magicity survives at the carbon drip line. 
The resulting interplay of weak binding, deformation, and shell evolution is correspondingly rich, yet existing descriptions have not converged on a single, convincing picture.

The halo observables themselves remain loosely constrained.
The reaction cross section on a hydrogen target gives an rms matter radius
$R_m=5.4\pm0.9$~fm~\cite{Tanaka2010_PRL104-062701}, whereas the interaction
cross section on a carbon target gives $3.44\pm0.08$~fm~\cite{Togano2016_PLB761-412};
a Glauber reanalysis of the latter data yields
$3.38\pm0.10$~fm~\cite{Nagahisa2018_PRC97-054614}.
The direct mass measurement gives $S_{2n}=-0.14\pm0.46$~MeV
\cite{Gaudefroy2012_PRL109-202503}, while the AME2020 value of
$0.035\pm0.020$~MeV is an evaluation rather than a new measurement
\cite{Wang2021_ChinPhysC45-030003}.
A narrow momentum distribution of the $^{20}$C residue after two-neutron
removal supports a two-neutron halo and is consistent with a substantial
$s$-wave component \cite{Kobayashi2012_PRC86-054604}, but it does not
determine the $d$-wave admixture.

Existing calculations convert these constraints into incompatible
structures.
Three-body models and halo effective field theory assume an inert
$^{20}$C core and two predominantly $s$-wave
neutrons~\cite{Horiuchi2006_PRC74-034311,Ershov2012_PRC86-034331,
Pinilla2016_PRC94-024620,Souza2016_PRC94-064002,
Souza2016_PLB757-368,Shulgina2018_PRC97-064307,
Acharya2013_PLB723-196,Hammer2017_JPG44-103002};
their radii and dipole strengths then hinge on $S_{2n}$ and on the
unknown $^{20}$C--$n$ scattering length.
Spherical density-functional calculations yield no pronounced
halo~\cite{Lu2013_PRC87-034311,Inakura2014_PRC89-064316}, 
whereas our earlier study~\cite{Sun2018_PLB785-530} with meson-exchange (ME)
functionals finds a shrunk, $sd$-mixed halo in the intrinsic
frame using the deformed relativistic Hartree--Bogoliubov theory in continuum (DRHBc) \cite{Zhou2010_PRC82-011301R}.
The only \textit{ab initio} calculation, a Gamow-IMSRG study, produces
an extended $s$-wave density without deformation~\cite{Hu2019_PRC99-061302R}.
Valence-space shell-model calculations~\cite{Coraggio2010_PRC81-064303,Yuan2012_PRC85-064324,Jansen:2014qxa}
and a beyond-mean-field (BMF) calculation~\cite{Yao2011_PRC84-024306}
already give laboratory-frame spectra of $^{22}$C, but both expand the
wave functions in a harmonic-oscillator basis and omit the continuum
required for a halo tail.

No existing description therefore combines a deformable,
pair-correlated $^{20}$C core, $s$-wave strength, and 
shell evolution in this mass region.
Without that combination one cannot decide whether $^{22}$C is an
inert-core $s$-wave halo, a deformed neighbor of $^{20}$C, or a
spherical $N=16$ system like $^{24}$O, nor compute a $B(E2)$ value that
distinguishes these pictures.
In this Letter, we develop a multi-reference density functional theory (MR-DFT) for deformed halo nuclei: particle number and angular
momentum are restored and quadrupole shapes are mixed, while the
continuum is retained.
We find a deformed two-neutron halo in $^{22}$C that is transitional
between deformed $^{20}$C and spherical $^{24}$O.
The valence neutrons occupy a mixed $sd$ configuration; the $N = 16$ closure of $^{24}$O is weakened in $^{22}$C.
A further result is that, with the point-coupling functionals used here, this
halo is absent at the mean-field (MF) level and is generated by the BMF correlations,
whereas meson-exchange functionals already produce it at the MF
state~\cite{Sun2018_PLB785-530}.
The deformed halo is thus robust in DFT studies.

{\it Theoretical framework.-}
The DRHBc theory \cite{Zhou2010_PRC82-011301R,Li2012_PRC85-024312} has been applied successfully to many halo systems 
(see reviews Refs.~\cite{Meng2015_JPG42-093101,Sun2024_NPR41-75,Zhang:2025cxi} 
and references therein)
and also to the construction of nuclear mass tables \cite{DRHBcMassTable:2022uhi,Guo2024_ADNDT158-101661}.
In this work, we implement the MR-DFT built on the DRHBc theory.
A single MF vacuum spontaneously breaks particle-number and
rotational invariance and contains no fluctuations about the average
deformation \cite{Ring1980,Egido2016_PS91-073003}.
These broken symmetries in intrinsically deformed
Bogoliubov states $\lvert\Phi(\beta)\rangle$ are restored 
using angular momentum
and particle number projections (AMP+PNP) after variation \cite{Ring1980}, and the missing fluctuations are recovered by mixing the projected configurations along
the quadrupole deformation 
\cite{Niksic2006_PRC73-034308,
Yao2010_PRC81-044311,Niksic2011_PPNP66-519} 
using the generator coordinate method (GCM),
\begin{equation}
\lvert\Psi^{JNZ}_\alpha\rangle
=\sum_{\beta}f^{JNZ}_\alpha(\beta)\,
\hat{P}^J\hat{P}^N\hat{P}^Z\lvert\Phi(\beta)\rangle.
\label{eq:mr-dft}
\end{equation}
For a weakly bound halo, the reference states $\lvert\Phi(\beta)\rangle$
cannot be ordinary MF vacua: they must already
couple bound orbitals to the continuum.
We therefore generate $\lvert\Phi(\beta)\rangle$ with the DRHBc theory
\cite{Zhou2010_PRC82-011301R,Li2012_PRC85-024312,Sun2018_PLB785-530}
and evaluate Eq.~\eqref{eq:mr-dft} while retaining the Dirac
Woods--Saxon continuum representation \cite{Zhou2003_PRC68-034323}, 
thereby extending the AMP of deformed continuum states
in Refs.~\cite{Sun2021_SciBull66-1521,Sun2021_PRC104-064319} 
to good particle number and full configuration mixing.
The Hill--Wheeler--Griffin amplitudes $f^{JNZ}_\alpha$ produce states of
good $J$, from which the spectrum, the $B(E2)$ values, and the underlying configurations are obtained. 
The collective wave functions $g_\alpha = N^{1/2} f_\alpha$, with $N$ the norm kernel, give the configuration content.
In the following, we refer to this new approach as MR-DRHBc.

Within covariant density functional theory, these BMF methods
are usually based on point-coupling interactions
\cite{Niksic2011_PPNP66-519}.
We use three functionals, PC-PK1 \cite{Zhao2010_PRC82-054319},
PC-F1 \cite{Burvenich2002_PRC65-044308}, and
DD-PC1 \cite{Niksic2008_PRC78-034318},
so that our conclusions do not depend on a single effective interaction.
The numerical details of the DRHBc calculations and of the AMP follow Refs.~\cite{Zhang2020_PRC102-024314,Sun2021_PRC104-064319}. 
The overlaps are calculated using the Pfaffian method \cite{Robledo2009_PRC79-021302} and the integral
over the gauge angles is calculated using Fomenko's method \cite{Fomenko1970_JPA3-8}.
The number of gauge angles 
is taken to be $7$ \cite{Yao2022_HNP,Zhou2023_IJMPE32-2340011}, for which we find no numerical instability. 
The AMP and PNP are performed
in the full canonical basis of each DRHBc vacuum, 
so that the discretized continuum already present at
the MF level is kept in the BMF level.
A zero-range density-dependent pairing force \cite{Meng1998_NPA635-3}
with a sharp energy
cutoff of $100$~MeV is employed in the particle-particle channel.
More details are given in the Sec.~\ref{sec:supp1}.

{\it Results and discussion.-} 
At the MF level, all three functionals give a spherical 
$^{22}$C with $sd$-orbital nearly degenerate, leading to a
energy gap at $N=16$, and the matter radius is about 3.11 fm, which is similar to
the cases in Refs. \cite{Sun2018_PLB785-530,Sun2020_NPA1003-122011} and donot support a halo.
Using MR-DRHBc, $R_{m}$ of $^{22}$C is enlarged to be 3.29 fm, 
about 0.17 fm above the $R_{m}$ of $^{20}$C, and closer to the interaction-cross-section extraction 
$3.44\pm0.08$ fm \cite{Togano2016_PLB761-412} and 
$3.38\pm0.10$ fm \cite{Nagahisa2018_PRC97-054614}.
At the MF level PC-F1 and PC-PK1 overbind $^{22}$C, giving
$S_{2n}\approx2$-3~MeV, while the BMF correlations reduce it to
$S_{2n}\approx0.6$--$0.8$~MeV, the range expected for a halo.
DD-PC1 gives a weakly bound $^{22}$C at the MF level and drives it
slightly unbound once the BMF correlation energy is included; its reference
states remain bound, so the GCM densities can still be evaluated in the finite
box. 
PC-F1 and PC-PK1 were adjusted to spherical finite nuclei including
light systems such as $^{16}$O and therefore
overbind $^{22}$C already at the MF level, whereas DD-PC1 was
fitted only to masses of heavy axially deformed nuclei and is not
constrained at $A\sim22$, so that the similar BMF correlation energy
unbinds it.
The expectation value and fluctuation of deformation
of GCM $0_{1}^{+}$ indicate that oblate deformations dominate it.
We have verified that the results in this work are insensitive to the box size.
More details of the bulk properties are 
provided in the Sec.~\ref{sec:bulk}.
MF ME functionals PK1 \cite{Long2004_PRC69-034319} 
and TMA \cite{Toki1995_NPA588-c357} calculations
already produce an oblate halo at the shallow global minimum of
the potential-energy curve
\cite{Sun2018_PLB785-530,Sun2020_NPA1003-122011}. 
That minimum is close to the same deformation that
dominates the present GCM $0^+_1$ weight with $|g(\beta)|^2$ peaking near
$\beta\approx-0.4$. 
Restoring symmetries and mixing shapes on those
references with ME functionals are therefore expected to broaden the collective wave function
about a deformed halo configuration.
What is specific to the point-coupling functionals used in this Letter is
that the halo is absent at the MF level and appears only after
configuration mixing.

\begin{figure}[ht]
  \centering
  \includegraphics[width=0.9\columnwidth]{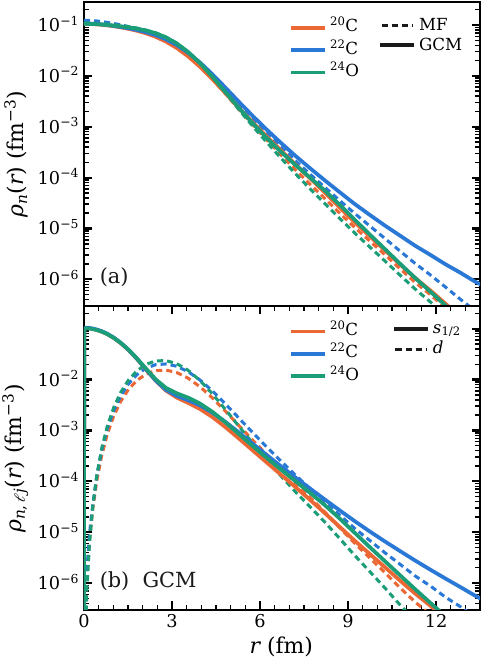}
\caption{Neutron densities of $^{20,22}$C and $^{24}$O
obtained with PC-F1.
(a)~Spherical monopole density $\rho_n(r)$ for the MF ground state (dashed lines) and for the $0^+_1$ state (solid lines).
(b)~Partial-wave neutron densities $\rho_{n,\ell j}(r)$ in the $0^+_1$ states: the
$s_{1/2}$ component (solid lines) and the $d_{5/2}$ and $d_{3/2}$ components (dashed lines, labeled ``$d$'').}
  \label{fig:den}
\end{figure}

One of the typical features of a nuclear halo is its large spatial extension.
Figure~\ref{fig:den}(a) shows the neutron densities obtained with PC-F1 for 
$^{20,22}$C and $^{24}$O, 
because the three density functionals give the same picture.
The neutron density of $^{22}$C has a long tail compared with
the other two, showing the halo nature, while the MF calculations do not
show this feature. 
In Fig. \ref{fig:den}(b), we show the partial wave decomposition of neutron density. 
The contributions from the neutron $1d_{{5/2}}$ and $1d_{{3/2}}$ (labeled as ``$d$'')
and $2s_{{1/2}}$ are given for the GCM ground states of three systems.
In $^{22}$C the $d$ components remain comparable to the $s_{1/2}$ one out to
$r\approx8$~fm, beyond which the $s_{1/2}$ component dominates, as expected from
the centrifugal suppression of the $d$ waves.
The configuration is therefore $sd$ mixed while the asymptotics are
$s$-dominated. 
The results from PC-PK1 and DD-PC1 are shown
in Sec.~\ref{sec:supp-den} and the conculsion 
is the same as those with PC-F1.

\begin{figure}[t]
  \centering
  \includegraphics[width=\columnwidth]{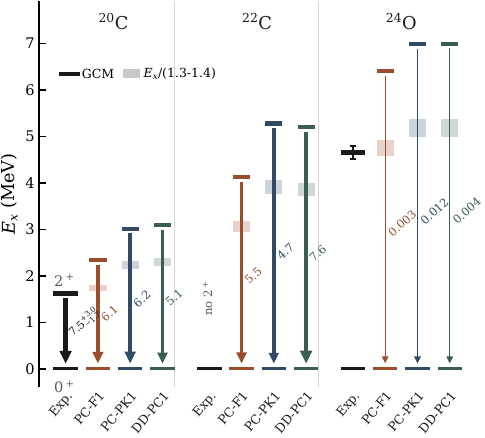}
  \caption{Low-lying $0^+_1$ and $2^+_1$ states of $^{20,22}$C and $^{24}$O, and $B(E2;2^+_1\to0^+_1)$ values in
    $e^2$fm$^4$ (numbers beside the arrows; the arrow width is
    scaled to the $B(E2)$). The shaded bars mark the interval
    $E_x/1.4$--$E_x/1.3$ as a visual guide to the systematic
    overestimate of the $2^+$ energies. 
    The experimental values are taken from Ref.~\cite{Petri2011_PRL107-102501} for $^{20}$C and from Ref.~\cite{Tshoo2012_PRL109-22501} for $^{24}$O.}
  \label{fig:spectra}
\end{figure}

\begin{figure}[t]
  \centering
  \includegraphics[width=\columnwidth]{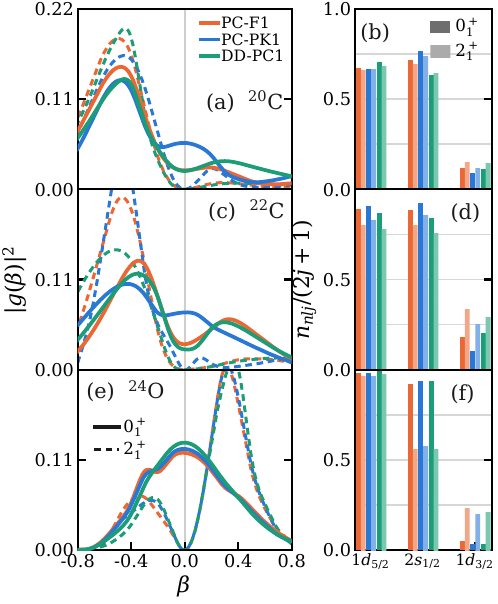}
\caption{GCM collective weights and neutron $sd$ occupations of
$^{20,22}$C and $^{24}$O.
(a),(c),(e)~Configuration content $|g(\beta)|^2$ of the yrast $0^+_1$
(solid lines) and $2^+_1$ (dotted lines) states along the axial quadrupole generator
coordinate. 
(b),(d),(f)~Woods--Saxon spherical occupations of the same GCM states,
shown as filling fractions $n_{nlj}/(2j+1)$, with
filled bars for $0^+_1$ and hatched bars for $2^+_1$.}
  \label{fig:g2-spl}
\end{figure}
Figure~\ref{fig:spectra} presents the low-lying spectra and
$B(E2;2^+_1\!\to\!0^+_1)$ values of $^{20,22}$C, and $^{24}$O 
obtained from the MR-DRHBc calculations compared with available data, 
together with the corresponding collective wave functions
$|g(\beta)|^2$ in Fig.~\ref{fig:g2-spl}.
The MF and projected potential energy curves 
are provided in Sec. \ref{sec:supp-den}.
The framework reproduces the established double magicity of
$^{24}$O: all three functionals place the $2^+_1$ state close to the measured
data~\cite{Tshoo2012_PRL109-22501}, yield very small $B(E2)$ values,
and give a $0^+_1$ collective wave function centered at sphericity.
The $2^{+}_{1}$ energy and $B(E2)$ for $^{20}$C's are also fairly
consistent with data.
Turning to $^{22}$C, we find a
qualitatively different situation.
Its $B(E2)$ values exceed those of $^{24}$O 
by two to three orders of magnitude and are comparable to those of $^{20}$C
($5.1$--$6.2$~$e^2$fm$^4$), and its $0^+_1$ collective wave function peaks at an
oblate deformation of $\beta\approx-0.4$, closely resembling that of $^{20}$C
rather than the spherical distribution of $^{24}$O.
The $N=16$ gap does, however, leave a clear imprint on the excitation energy:
$E(2^+_1)$ of $^{22}$C is roughly 1.5 times that of $^{20}$C for all three functionals.
Since the collective wave functions are similar between the two
isotopes, indicating a reduction of the collective moment of inertia
at essentially fixed deformation, rather than a restoration of sphericity.
We note that the calculated excitation energies are systematically overestimated,
by about $30$--$40\%$ for $^{20}$C, where experimental data are available; this is
a known consequence of the moment of inertia being underestimated by angular-momentum
projection~\cite{Niksic2007_PRL99-092502,Niksic2011_PPNP66-519}. 
The
shaded boxes in Fig.~\ref{fig:spectra} indicate the energies scaled by this
factor, and we base our conclusions primarily on ratios, which are insensitive to
this systematic shift.
Finally, we point out that, while all functionals agree on the large collectivity
of $^{22}$C relative to $^{24}$O, they differ on the detailed evolution of
$B(E2)$ from $^{20}$C to $^{22}$C. 
PC-PK1 yields a flatter, nearly
bimodal collective wave function for $^{22}$C.
With PC-F1 the collectivity of $^{20}$C agrees with
Yao~\textit{et~al.}~\cite{Yao2011_PRC84-024306}; their oscillator-basis
wave functions omit the continuum required for the halo tail of $^{22}$C.

The right panel of Fig.~\ref{fig:g2-spl} shows
the configuration in the $sd$ shell for GCM states, 
through the occupations $n_{nlj}$ of the
spherical Dirac Woods-Saxon orbitals shown as filling fractions
$n_{nlj}/(2j+1)$.
In the $0^+_1$ state of $^{24}$O, the $N=16$ closure is clearly manifested: the $1d_{5/2}$ and
$2s_{1/2}$ orbitals are almost fully filled, while the
occupation of
$1d_{3/2}$ orbital is tiny, being consistent with
the collective wave function centered at $\beta\approx0$.
In $^{20}$C, the $0^+_1$ wave function peaks 
at $\beta\approx-0.4$ and the $1d_{5/2}$ filling is 
$0.67$--$0.70$, whereas the $2s_{1/2}$ orbital already carries $0.63$--$0.77$
and the $1d_{3/2}$ orbital $0.09$--$0.12$,
reflecting the mixing of $sd$ orbitals due to deformation, thus
leading to the disappearance of subshell closure at $N=14$ \cite{Stanoiu2008_PRC78-034315}.
In $^{22}$C, the two mechanisms act together: the occupations of $sd$ states lie between the two limits.
The $1d_{5/2}$ filling, $0.87$--$0.92$, is reduced with respect to the closed
subshell of $^{24}$O but far above the strongly mixed value of $^{20}$C, the
$2s_{1/2}$ orbital carries $0.84$--$0.93$, and the $1d_{3/2}$ orbital takes up
$0.10$--$0.20$, corresponding to $0.4$--$0.8$ neutrons across the $N=16$ gap,
several times more than in $^{24}$O.
Therefore,
the $N = 16$ shell closure established in $^{24}$O is thus substantially weakened, though not removed, in $^{22}$C.
The increased occupation of the $2s_{1/2}$ orbital from $^{20}$C to $^{22}$C 
generates the halo seen in Fig. \ref{fig:den} with the valence neutrons remaining in an $s$-$d$ mixture.

The $0^+_1$ and $2^+_1$ occupations follow the same
hierarchy and identify the character of the excitation
as shown in Fig. \ref{fig:g2-spl}.
In $^{24}$O the $2^+_1$ state is reached by nearly moving about $0.7$ neutrons from
$2s_{1/2}$ into $1d_{3/2}$: a particle-hole excitation across the
$N=16$ gap rather than a rotation, which is precisely why its $B(E2)$ value is
vanishingly small.
In $^{20}$C the two sets of occupations are nearly identical, which is a rotational partner built on the
same intrinsic configuration.
$^{22}$C lies in between, with occupation changes of about $0.1$ and a
$1d_{3/2}$ filling that rises from $0.10$--$0.20$ to $0.29$--$0.33$: its $2^+_1$
state is predominantly rotational, but it acquires a visible single-particle
component.
PC-PK1 gives the smallest $1d_{3/2}$ occupation in the $0^+_1$ state
of $^{22}$C.  
Its $|g(\beta)|^2$ is correspondingly the flattest of the three, with
$\langle\beta\rangle=-0.19$ for the $0^+_1$ state and $-0.4$ for the $2^+_1$
state.
Even in this most $N=16$-closure-like case, however, the broad shape mixing and the
still-finite $B(E2)$ describe a shape-soft, $sd$-mixed structure rather than a
rigid $N=16$ closure found in $^{24}$O.
The $0_{1}^{+}$ state of $^{22}$C also carries a small $p$-wave occupation number about 0.05, originating from $2p$ orbitals embedded in the continuum.

\begin{figure}[ht]
  \centering
  \includegraphics[width=0.9\columnwidth]{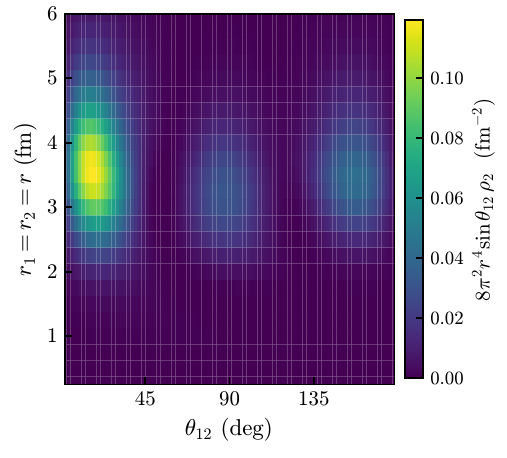}
  \caption{Spin-singlet $nn$ correlation density $\rho_2$
  of $^{22}$C, evaluated for $J^{\pi}=0^{+}$ projected generator at $\beta_{2}=-0.4$.
  }
  \label{fig:corr}
\end{figure}

The halo configurations can form unique neutron-neutron ($nn$) correlations.
The spin-singlet $nn$ correlation density $\rho_2$ 
is displayed versus $(r,\theta_{12})$ at $r_1=r_2=r$ as
$8\pi^2 r^4\sin\theta_{12}\,\rho_2$ following 
Refs.~\cite{Hagino2005_PRC72-44321,Matsuo2005_PRC71-064326} and 
the detail of the calcuations is shown Sec.~\ref{sec:supp-p2}.
We mention that {\it ab initio} theories
also explore such correlations, but only in lighter systems \cite{Navratil2026_Particle9-57,Zhang2026,Yang2026}.
We evaluate $\rho_2$ for the projected $0^+$ generator at
$\beta=-0.4$, the oblate peak of $|g(\beta)|^2$ in the
$0^+_1$ state.
Starting from the canonical vacuum of this generator, the
laboratory-frame $J^\pi=0^+$ map peaks at $r=3.5$~fm and
$\theta_{12}=18^\circ$ as shown in Fig. \ref{fig:corr} using PC-F1: 
a dineutron configuration together with a weak cigar-like one
at $\theta_{12}\sim 180^\circ$.
Similar results are also obtained for PC-PK1 and DD-PC1 (see Sec.~\ref{sec:supp-p2}).
This dineutron localization arises from the coherent superposition
of opposite-parity $sd$ and $p$ components~\cite{Hagino2005_PRC72-44321}.
After integrating over radius, about $59\%$ of the strength lies at
$\theta_{12}<60^\circ$ and only $\sim 22\%$ at $\theta_{12}>120^\circ$.
This reflects that
the weakly bound valence pair in $^{22}$C moves together relative to the
core.
The same geometry fixes the soft $E1$ strength through the cluster sum
rule
$B(E1)=(3/4\pi)e^{2}(2Z/A)^{2}\langle r_{c\text{-}2n}^{2}\rangle$
\cite{Alhassid1982_PRL49-1482} with
$r_{c\text{-}2n}$ being the distance between the pair of neutrons 
to the center of the core.
Folding the GCM radii of $^{22}$C and $^{20}$C into this dineutron limit
gives $B(E1)\sim 1.5\,e^{2}\mathrm{fm}^{2}$, an integrated strength, not
a dipole response spectrum; it can be tested by forthcoming Coulomb
dissociation data.
The same dineutron geometry is obtained for generators throughout the oblate
region that dominates $|g(\beta)|^2$, so this correlation is not specific to a
single reference state.

{\it Summary.-} 
We develop the MR-DRHBc approach to 
achieve self-consistent descriptions for the collective motion of deformed exotic nuclei.
This new approach provides a unified description of $^{22}$C: a deformed
two-neutron halo with $R_m \approx 3.29\ \mathrm{fm}$, closer to the
carbon-target extractions, a long $s$-wave tail, and dineutron-dominated
pair correlations from mixed $sd$ valence neutrons with a small $p$-wave
admixture.
It simultaneously accounts for the spectra and $B(E2)$ values of
$^{20}$C, $^{22}$C, and $^{24}$O, placing $^{22}$C in the transition
between deformed $^{20}$C and spherical doubly magic $^{24}$O.
Radius, collectivity, shell evolution, and spatial pair correlations
are obtained in a self-consistent framework.
The same framework applies to deformed halo nuclei more generally and,
once extended to odd-$A$ systems, can treat deformed halo systems on
the same footing.

\begin{acknowledgments}
The author thanks helpful discussions with Shan-Gui Zhou and Ulf-G. Mei{\ss}ner.
This work was supported in part by the National Natural Science Foundation of China under Grant No.~12205308.
The numerical calculations were performed on JURECA
of the J\"ulich Supercomputing Center, J\"ulich, Germany, and 
on the ITP-CAS High-performance Computing Cluster of ITP-CAS and the ScGrid of the Supercomputing Center, Computer Network Information Center of the Chinese Academy of Sciences. 
AI assistants (Cursor, Grok~4.6; Claude) were only used to help
revise the English wording of the scientific claims and to speed up loop-level implementations in the
production code.

\end{acknowledgments}

\begin{thebibliography}{72}%
\makeatletter
\providecommand \@ifxundefined [1]{%
 \@ifx{#1\undefined}
}%
\providecommand \@ifnum [1]{%
 \ifnum #1\expandafter \@firstoftwo
 \else \expandafter \@secondoftwo
 \fi
}%
\providecommand \@ifx [1]{%
 \ifx #1\expandafter \@firstoftwo
 \else \expandafter \@secondoftwo
 \fi
}%
\providecommand \natexlab [1]{#1}%
\providecommand \enquote  [1]{``#1''}%
\providecommand \bibnamefont  [1]{#1}%
\providecommand \bibfnamefont [1]{#1}%
\providecommand \citenamefont [1]{#1}%
\providecommand \href@noop [0]{\@secondoftwo}%
\providecommand \href [0]{\begingroup \@sanitize@url \@href}%
\providecommand \@href[1]{\@@startlink{#1}\@@href}%
\providecommand \@@href[1]{\endgroup#1\@@endlink}%
\providecommand \@sanitize@url [0]{\catcode `\\12\catcode `\$12\catcode
  `\&12\catcode `\#12\catcode `\^12\catcode `\_12\catcode `\%12\relax}%
\providecommand \@@startlink[1]{}%
\providecommand \@@endlink[0]{}%
\providecommand \url  [0]{\begingroup\@sanitize@url \@url }%
\providecommand \@url [1]{\endgroup\@href {#1}{\urlprefix }}%
\providecommand \urlprefix  [0]{URL }%
\providecommand \Eprint [0]{\href }%
\providecommand \doibase [0]{https://doi.org/}%
\providecommand \selectlanguage [0]{\@gobble}%
\providecommand \bibinfo  [0]{\@secondoftwo}%
\providecommand \bibfield  [0]{\@secondoftwo}%
\providecommand \translation [1]{[#1]}%
\providecommand \BibitemOpen [0]{}%
\providecommand \bibitemStop [0]{}%
\providecommand \bibitemNoStop [0]{.\EOS\space}%
\providecommand \EOS [0]{\spacefactor3000\relax}%
\providecommand \BibitemShut  [1]{\csname bibitem#1\endcsname}%
\let\auto@bib@innerbib\@empty
\bibitem [{\citenamefont {Jensen}\ \emph {et~al.}(2004)\citenamefont {Jensen},
  \citenamefont {Riisager}, \citenamefont {Fedorov},\ and\ \citenamefont
  {Garrido}}]{Jensen2004_RMP76-215}%
  \BibitemOpen
  \bibfield  {author} {\bibinfo {author} {\bibfnamefont {A.~S.}\ \bibnamefont
  {Jensen}}, \bibinfo {author} {\bibfnamefont {K.}~\bibnamefont {Riisager}},
  \bibinfo {author} {\bibfnamefont {D.~V.}\ \bibnamefont {Fedorov}},\ and\
  \bibinfo {author} {\bibfnamefont {E.}~\bibnamefont {Garrido}},\ }\href
  {https://doi.org/10.1103/RevModPhys.76.215} {\bibfield  {journal} {\bibinfo
  {journal} {Rev. Mod. Phys.}\ }\textbf {\bibinfo {volume} {76}},\ \bibinfo
  {pages} {215} (\bibinfo {year} {2004})}\BibitemShut {NoStop}%
\bibitem [{\citenamefont {Tanihata}\ \emph {et~al.}(2013)\citenamefont
  {Tanihata}, \citenamefont {Savajols},\ and\ \citenamefont
  {Kanungo}}]{Tanihata2013_PPNP68-215}%
  \BibitemOpen
  \bibfield  {author} {\bibinfo {author} {\bibfnamefont {I.}~\bibnamefont
  {Tanihata}}, \bibinfo {author} {\bibfnamefont {H.}~\bibnamefont {Savajols}},\
  and\ \bibinfo {author} {\bibfnamefont {R.}~\bibnamefont {Kanungo}},\ }\href
  {https://doi.org/10.1016/j.ppnp.2012.07.001} {\bibfield  {journal} {\bibinfo
  {journal} {Prog. Part. Nucl. Phys.}\ }\textbf {\bibinfo {volume} {68}},\
  \bibinfo {pages} {215} (\bibinfo {year} {2013})}\BibitemShut {NoStop}%
\bibitem [{\citenamefont {Meng}\ and\ \citenamefont
  {Zhou}(2015)}]{Meng2015_JPG42-093101}%
  \BibitemOpen
  \bibfield  {author} {\bibinfo {author} {\bibfnamefont {J.}~\bibnamefont
  {Meng}}\ and\ \bibinfo {author} {\bibfnamefont {S.-G.}\ \bibnamefont
  {Zhou}},\ }\href {https://doi.org/10.1088/0954-3899/42/9/093101} {\bibfield
  {journal} {\bibinfo  {journal} {J. Phys. G: Nucl. Part. Phys.}\ }\textbf
  {\bibinfo {volume} {42}},\ \bibinfo {pages} {093101} (\bibinfo {year}
  {2015})}\BibitemShut {NoStop}%
\bibitem [{\citenamefont {Nakamura}(2019)}]{Nakamura2019_AAPPSBL29-5}%
  \BibitemOpen
  \bibfield  {author} {\bibinfo {author} {\bibfnamefont {T.}~\bibnamefont
  {Nakamura}},\ }\bibfield  {journal} {\bibinfo  {journal} {AAPPS Bulletin}\
  }\textbf {\bibinfo {volume} {29}},\ \href
  {https://doi.org/10.22661/AAPPSBL.2019.29.5.19}
  {10.22661/AAPPSBL.2019.29.5.19} (\bibinfo {year} {2019})\BibitemShut
  {NoStop}%
\bibitem [{\citenamefont {Misu}\ \emph {et~al.}(1997)\citenamefont {Misu},
  \citenamefont {Nazarewicz},\ and\ \citenamefont
  {{\.A}berg}}]{Misu1997_NPA614-44}%
  \BibitemOpen
  \bibfield  {author} {\bibinfo {author} {\bibfnamefont {T.}~\bibnamefont
  {Misu}}, \bibinfo {author} {\bibfnamefont {W.}~\bibnamefont {Nazarewicz}},\
  and\ \bibinfo {author} {\bibfnamefont {S.}~\bibnamefont {{\.A}berg}},\ }\href
  {https://doi.org/10.1016/S0375-9474(96)00458-7} {\bibfield  {journal}
  {\bibinfo  {journal} {Nucl. Phys. A}\ }\textbf {\bibinfo {volume} {614}},\
  \bibinfo {pages} {44} (\bibinfo {year} {1997})}\BibitemShut {NoStop}%
\bibitem [{\citenamefont {Hamamoto}(2010)}]{Hamamoto2010_PRC81-021304}%
  \BibitemOpen
  \bibfield  {author} {\bibinfo {author} {\bibfnamefont {I.}~\bibnamefont
  {Hamamoto}},\ }\href {https://doi.org/10.1103/PhysRevC.81.021304} {\bibfield
  {journal} {\bibinfo  {journal} {Phys. Rev. C}\ }\textbf {\bibinfo {volume}
  {81}},\ \bibinfo {pages} {021304(R)} (\bibinfo {year} {2010})}\BibitemShut
  {NoStop}%
\bibitem [{\citenamefont {Zhou}\ \emph {et~al.}(2010)\citenamefont {Zhou},
  \citenamefont {Meng}, \citenamefont {Ring},\ and\ \citenamefont
  {Zhao}}]{Zhou2010_PRC82-011301R}%
  \BibitemOpen
  \bibfield  {author} {\bibinfo {author} {\bibfnamefont {S.-G.}\ \bibnamefont
  {Zhou}}, \bibinfo {author} {\bibfnamefont {J.}~\bibnamefont {Meng}}, \bibinfo
  {author} {\bibfnamefont {P.}~\bibnamefont {Ring}},\ and\ \bibinfo {author}
  {\bibfnamefont {E.-G.}\ \bibnamefont {Zhao}},\ }\href
  {https://doi.org/10.1103/physrevc.82.011301} {\bibfield  {journal} {\bibinfo
  {journal} {Phys. Rev. C}\ }\textbf {\bibinfo {volume} {82}},\ \bibinfo
  {pages} {011301(R)} (\bibinfo {year} {2010})}\BibitemShut {NoStop}%
\bibitem [{\citenamefont {Nakamura}\ \emph {et~al.}(2009)\citenamefont
  {Nakamura}, \citenamefont {Kobayashi}, \citenamefont {Kondo}, \citenamefont
  {Satou}, \citenamefont {Aoi}, \citenamefont {Baba}, \citenamefont {Deguchi},
  \citenamefont {Fukuda}, \citenamefont {Gibelin}, \citenamefont {Inabe},
  \citenamefont {Ishihara}, \citenamefont {Kameda}, \citenamefont {Kawada},
  \citenamefont {Kubo}, \citenamefont {Kusaka}, \citenamefont {Mengoni},
  \citenamefont {Motobayashi}, \citenamefont {Ohnishi}, \citenamefont {Ohtake},
  \citenamefont {Orr}, \citenamefont {Otsu}, \citenamefont {Otsuka},
  \citenamefont {Saito}, \citenamefont {Sakurai}, \citenamefont {Shimoura},
  \citenamefont {Sumikama}, \citenamefont {Takeda}, \citenamefont {Takeshita},
  \citenamefont {Takechi}, \citenamefont {Takeuchi}, \citenamefont {Tanaka},
  \citenamefont {Tanaka}, \citenamefont {Tanaka}, \citenamefont {Togano},
  \citenamefont {Utsuno}, \citenamefont {Yoneda}, \citenamefont {Yoshida},\
  and\ \citenamefont {Yoshida}}]{Nakamura2009_PRL103-262501}%
  \BibitemOpen
  \bibfield  {author} {\bibinfo {author} {\bibfnamefont {T.}~\bibnamefont
  {Nakamura}}, \bibinfo {author} {\bibfnamefont {N.}~\bibnamefont {Kobayashi}},
  \bibinfo {author} {\bibfnamefont {Y.}~\bibnamefont {Kondo}}, \bibinfo
  {author} {\bibfnamefont {Y.}~\bibnamefont {Satou}}, \bibinfo {author}
  {\bibfnamefont {N.}~\bibnamefont {Aoi}}, \bibinfo {author} {\bibfnamefont
  {H.}~\bibnamefont {Baba}}, \bibinfo {author} {\bibfnamefont {S.}~\bibnamefont
  {Deguchi}}, \bibinfo {author} {\bibfnamefont {N.}~\bibnamefont {Fukuda}},
  \bibinfo {author} {\bibfnamefont {J.}~\bibnamefont {Gibelin}}, \bibinfo
  {author} {\bibfnamefont {N.}~\bibnamefont {Inabe}}, \bibinfo {author}
  {\bibfnamefont {M.}~\bibnamefont {Ishihara}}, \bibinfo {author}
  {\bibfnamefont {D.}~\bibnamefont {Kameda}}, \bibinfo {author} {\bibfnamefont
  {Y.}~\bibnamefont {Kawada}}, \bibinfo {author} {\bibfnamefont
  {T.}~\bibnamefont {Kubo}}, \bibinfo {author} {\bibfnamefont {K.}~\bibnamefont
  {Kusaka}}, \bibinfo {author} {\bibfnamefont {A.}~\bibnamefont {Mengoni}},
  \bibinfo {author} {\bibfnamefont {T.}~\bibnamefont {Motobayashi}}, \bibinfo
  {author} {\bibfnamefont {T.}~\bibnamefont {Ohnishi}}, \bibinfo {author}
  {\bibfnamefont {M.}~\bibnamefont {Ohtake}}, \bibinfo {author} {\bibfnamefont
  {N.~A.}\ \bibnamefont {Orr}}, \bibinfo {author} {\bibfnamefont
  {H.}~\bibnamefont {Otsu}}, \bibinfo {author} {\bibfnamefont {T.}~\bibnamefont
  {Otsuka}}, \bibinfo {author} {\bibfnamefont {A.}~\bibnamefont {Saito}},
  \bibinfo {author} {\bibfnamefont {H.}~\bibnamefont {Sakurai}}, \bibinfo
  {author} {\bibfnamefont {S.}~\bibnamefont {Shimoura}}, \bibinfo {author}
  {\bibfnamefont {T.}~\bibnamefont {Sumikama}}, \bibinfo {author}
  {\bibfnamefont {H.}~\bibnamefont {Takeda}}, \bibinfo {author} {\bibfnamefont
  {E.}~\bibnamefont {Takeshita}}, \bibinfo {author} {\bibfnamefont
  {M.}~\bibnamefont {Takechi}}, \bibinfo {author} {\bibfnamefont
  {S.}~\bibnamefont {Takeuchi}}, \bibinfo {author} {\bibfnamefont
  {K.}~\bibnamefont {Tanaka}}, \bibinfo {author} {\bibfnamefont {K.~N.}\
  \bibnamefont {Tanaka}}, \bibinfo {author} {\bibfnamefont {N.}~\bibnamefont
  {Tanaka}}, \bibinfo {author} {\bibfnamefont {Y.}~\bibnamefont {Togano}},
  \bibinfo {author} {\bibfnamefont {Y.}~\bibnamefont {Utsuno}}, \bibinfo
  {author} {\bibfnamefont {K.}~\bibnamefont {Yoneda}}, \bibinfo {author}
  {\bibfnamefont {A.}~\bibnamefont {Yoshida}},\ and\ \bibinfo {author}
  {\bibfnamefont {K.}~\bibnamefont {Yoshida}},\ }\href
  {https://doi.org/10.1103/PhysRevLett.103.262501} {\bibfield  {journal}
  {\bibinfo  {journal} {Phys. Rev. Lett.}\ }\textbf {\bibinfo {volume} {103}},\
  \bibinfo {pages} {262501} (\bibinfo {year} {2009})}\BibitemShut {NoStop}%
\bibitem [{\citenamefont {Nakamura}\ \emph {et~al.}(2014)\citenamefont
  {Nakamura}, \citenamefont {Kobayashi}, \citenamefont {Kondo}, \citenamefont
  {Satou}, \citenamefont {Tostevin}, \citenamefont {Utsuno}, \citenamefont
  {Aoi}, \citenamefont {Baba}, \citenamefont {Fukuda}, \citenamefont {Gibelin},
  \citenamefont {Inabe}, \citenamefont {Ishihara}, \citenamefont {Kameda},
  \citenamefont {Kubo}, \citenamefont {Motobayashi}, \citenamefont {Ohnishi},
  \citenamefont {Orr}, \citenamefont {Otsu}, \citenamefont {Otsuka},
  \citenamefont {Sakurai}, \citenamefont {Sumikama}, \citenamefont {Takeda},
  \citenamefont {Takeshita}, \citenamefont {Takechi}, \citenamefont {Takeuchi},
  \citenamefont {Togano},\ and\ \citenamefont
  {Yoneda}}]{Nakamura2014_PRL112-142501}%
  \BibitemOpen
  \bibfield  {author} {\bibinfo {author} {\bibfnamefont {T.}~\bibnamefont
  {Nakamura}}, \bibinfo {author} {\bibfnamefont {N.}~\bibnamefont {Kobayashi}},
  \bibinfo {author} {\bibfnamefont {Y.}~\bibnamefont {Kondo}}, \bibinfo
  {author} {\bibfnamefont {Y.}~\bibnamefont {Satou}}, \bibinfo {author}
  {\bibfnamefont {J.~A.}\ \bibnamefont {Tostevin}}, \bibinfo {author}
  {\bibfnamefont {Y.}~\bibnamefont {Utsuno}}, \bibinfo {author} {\bibfnamefont
  {N.}~\bibnamefont {Aoi}}, \bibinfo {author} {\bibfnamefont {H.}~\bibnamefont
  {Baba}}, \bibinfo {author} {\bibfnamefont {N.}~\bibnamefont {Fukuda}},
  \bibinfo {author} {\bibfnamefont {J.}~\bibnamefont {Gibelin}}, \bibinfo
  {author} {\bibfnamefont {N.}~\bibnamefont {Inabe}}, \bibinfo {author}
  {\bibfnamefont {M.}~\bibnamefont {Ishihara}}, \bibinfo {author}
  {\bibfnamefont {D.}~\bibnamefont {Kameda}}, \bibinfo {author} {\bibfnamefont
  {T.}~\bibnamefont {Kubo}}, \bibinfo {author} {\bibfnamefont {T.}~\bibnamefont
  {Motobayashi}}, \bibinfo {author} {\bibfnamefont {T.}~\bibnamefont
  {Ohnishi}}, \bibinfo {author} {\bibfnamefont {N.~A.}\ \bibnamefont {Orr}},
  \bibinfo {author} {\bibfnamefont {H.}~\bibnamefont {Otsu}}, \bibinfo {author}
  {\bibfnamefont {T.}~\bibnamefont {Otsuka}}, \bibinfo {author} {\bibfnamefont
  {H.}~\bibnamefont {Sakurai}}, \bibinfo {author} {\bibfnamefont
  {T.}~\bibnamefont {Sumikama}}, \bibinfo {author} {\bibfnamefont
  {H.}~\bibnamefont {Takeda}}, \bibinfo {author} {\bibfnamefont
  {E.}~\bibnamefont {Takeshita}}, \bibinfo {author} {\bibfnamefont
  {M.}~\bibnamefont {Takechi}}, \bibinfo {author} {\bibfnamefont
  {S.}~\bibnamefont {Takeuchi}}, \bibinfo {author} {\bibfnamefont
  {Y.}~\bibnamefont {Togano}},\ and\ \bibinfo {author} {\bibfnamefont
  {K.}~\bibnamefont {Yoneda}},\ }\href
  {https://doi.org/10.1103/PhysRevLett.112.142501} {\bibfield  {journal}
  {\bibinfo  {journal} {Phys. Rev. Lett.}\ }\textbf {\bibinfo {volume} {112}},\
  \bibinfo {pages} {142501} (\bibinfo {year} {2014})}\BibitemShut {NoStop}%
\bibitem [{\citenamefont {Kobayashi}\ \emph {et~al.}(2014)\citenamefont
  {Kobayashi}, \citenamefont {Nakamura}, \citenamefont {Kondo}, \citenamefont
  {Tostevin}, \citenamefont {Utsuno}, \citenamefont {Aoi}, \citenamefont
  {Baba}, \citenamefont {Barthelemy}, \citenamefont {Famiano}, \citenamefont
  {Fukuda}, \citenamefont {Inabe}, \citenamefont {Ishihara}, \citenamefont
  {Kanungo}, \citenamefont {Kim}, \citenamefont {Kubo}, \citenamefont {Lee},
  \citenamefont {Lee}, \citenamefont {Matsushita}, \citenamefont {Motobayashi},
  \citenamefont {Ohnishi}, \citenamefont {Orr}, \citenamefont {Otsu},
  \citenamefont {Otsuka}, \citenamefont {Sako}, \citenamefont {Sakurai},
  \citenamefont {Satou}, \citenamefont {Sumikama}, \citenamefont {Takeda},
  \citenamefont {Takeuchi}, \citenamefont {Tanaka}, \citenamefont {Togano},\
  and\ \citenamefont {Yoneda}}]{Kobayashi2014_PRL112-242501}%
  \BibitemOpen
  \bibfield  {author} {\bibinfo {author} {\bibfnamefont {N.}~\bibnamefont
  {Kobayashi}}, \bibinfo {author} {\bibfnamefont {T.}~\bibnamefont {Nakamura}},
  \bibinfo {author} {\bibfnamefont {Y.}~\bibnamefont {Kondo}}, \bibinfo
  {author} {\bibfnamefont {J.~A.}\ \bibnamefont {Tostevin}}, \bibinfo {author}
  {\bibfnamefont {Y.}~\bibnamefont {Utsuno}}, \bibinfo {author} {\bibfnamefont
  {N.}~\bibnamefont {Aoi}}, \bibinfo {author} {\bibfnamefont {H.}~\bibnamefont
  {Baba}}, \bibinfo {author} {\bibfnamefont {R.}~\bibnamefont {Barthelemy}},
  \bibinfo {author} {\bibfnamefont {M.~A.}\ \bibnamefont {Famiano}}, \bibinfo
  {author} {\bibfnamefont {N.}~\bibnamefont {Fukuda}}, \bibinfo {author}
  {\bibfnamefont {N.}~\bibnamefont {Inabe}}, \bibinfo {author} {\bibfnamefont
  {M.}~\bibnamefont {Ishihara}}, \bibinfo {author} {\bibfnamefont
  {R.}~\bibnamefont {Kanungo}}, \bibinfo {author} {\bibfnamefont
  {S.}~\bibnamefont {Kim}}, \bibinfo {author} {\bibfnamefont {T.}~\bibnamefont
  {Kubo}}, \bibinfo {author} {\bibfnamefont {G.~S.}\ \bibnamefont {Lee}},
  \bibinfo {author} {\bibfnamefont {H.~S.}\ \bibnamefont {Lee}}, \bibinfo
  {author} {\bibfnamefont {M.}~\bibnamefont {Matsushita}}, \bibinfo {author}
  {\bibfnamefont {T.}~\bibnamefont {Motobayashi}}, \bibinfo {author}
  {\bibfnamefont {T.}~\bibnamefont {Ohnishi}}, \bibinfo {author} {\bibfnamefont
  {N.~A.}\ \bibnamefont {Orr}}, \bibinfo {author} {\bibfnamefont
  {H.}~\bibnamefont {Otsu}}, \bibinfo {author} {\bibfnamefont {T.}~\bibnamefont
  {Otsuka}}, \bibinfo {author} {\bibfnamefont {T.}~\bibnamefont {Sako}},
  \bibinfo {author} {\bibfnamefont {H.}~\bibnamefont {Sakurai}}, \bibinfo
  {author} {\bibfnamefont {Y.}~\bibnamefont {Satou}}, \bibinfo {author}
  {\bibfnamefont {T.}~\bibnamefont {Sumikama}}, \bibinfo {author}
  {\bibfnamefont {H.}~\bibnamefont {Takeda}}, \bibinfo {author} {\bibfnamefont
  {S.}~\bibnamefont {Takeuchi}}, \bibinfo {author} {\bibfnamefont
  {R.}~\bibnamefont {Tanaka}}, \bibinfo {author} {\bibfnamefont
  {Y.}~\bibnamefont {Togano}},\ and\ \bibinfo {author} {\bibfnamefont
  {K.}~\bibnamefont {Yoneda}},\ }\href
  {https://doi.org/10.1103/PhysRevLett.112.242501} {\bibfield  {journal}
  {\bibinfo  {journal} {Phys. Rev. Lett.}\ }\textbf {\bibinfo {volume} {112}},\
  \bibinfo {pages} {242501} (\bibinfo {year} {2014})}\BibitemShut {NoStop}%
\bibitem [{\citenamefont {Yang}\ \emph {et~al.}(2021)\citenamefont {Yang},
  \citenamefont {Kubota}, \citenamefont {Corsi}, \citenamefont {Yoshida},
  \citenamefont {Sun}, \citenamefont {Li}, \citenamefont {Kimura},
  \citenamefont {Michel}, \citenamefont {Ogata}, \citenamefont {Yuan},
  \citenamefont {Yuan}, \citenamefont {Authelet}, \citenamefont {Baba},
  \citenamefont {Caesar}, \citenamefont {Calvet}, \citenamefont {Delbart},
  \citenamefont {Dozono}, \citenamefont {Feng}, \citenamefont {Flavigny},
  \citenamefont {Gheller}, \citenamefont {Gibelin}, \citenamefont {Giganon},
  \citenamefont {Gillibert}, \citenamefont {Hasegawa}, \citenamefont {Isobe},
  \citenamefont {Kanaya}, \citenamefont {Kawakami}, \citenamefont {Kim},
  \citenamefont {Kiyokawa}, \citenamefont {Kobayashi}, \citenamefont
  {Kobayashi}, \citenamefont {Kobayashi}, \citenamefont {Kondo}, \citenamefont
  {Korkulu}, \citenamefont {Koyama}, \citenamefont {Lapoux}, \citenamefont
  {Maeda}, \citenamefont {Marqu{\'e}s}, \citenamefont {Motobayashi},
  \citenamefont {Miyazaki}, \citenamefont {Nakamura}, \citenamefont
  {Nakatsuka}, \citenamefont {Nishio}, \citenamefont {Obertelli}, \citenamefont
  {Ohkura}, \citenamefont {Orr}, \citenamefont {Ota}, \citenamefont {Otsu},
  \citenamefont {Ozaki}, \citenamefont {Panin}, \citenamefont {Paschalis},
  \citenamefont {Pollacco}, \citenamefont {Reichert}, \citenamefont
  {Rouss{\'e}}, \citenamefont {Saito}, \citenamefont {Sakaguchi}, \citenamefont
  {Sako}, \citenamefont {Santamaria}, \citenamefont {Sasano}, \citenamefont
  {Sato}, \citenamefont {Shikata}, \citenamefont {Shimizu}, \citenamefont
  {Shindo}, \citenamefont {Stuhl}, \citenamefont {Sumikama}, \citenamefont
  {Sun}, \citenamefont {Tabata}, \citenamefont {Togano}, \citenamefont
  {Tsubota}, \citenamefont {Xu}, \citenamefont {Yasuda}, \citenamefont
  {Yoneda}, \citenamefont {Zenihiro}, \citenamefont {Zhou}, \citenamefont
  {Zuo},\ and\ \citenamefont {Uesaka}}]{Yang2021_PRL126-082501}%
  \BibitemOpen
  \bibfield  {author} {\bibinfo {author} {\bibfnamefont {Z.~H.}\ \bibnamefont
  {Yang}}, \bibinfo {author} {\bibfnamefont {Y.}~\bibnamefont {Kubota}},
  \bibinfo {author} {\bibfnamefont {A.}~\bibnamefont {Corsi}}, \bibinfo
  {author} {\bibfnamefont {K.}~\bibnamefont {Yoshida}}, \bibinfo {author}
  {\bibfnamefont {X.-X.}\ \bibnamefont {Sun}}, \bibinfo {author} {\bibfnamefont
  {J.~G.}\ \bibnamefont {Li}}, \bibinfo {author} {\bibfnamefont
  {M.}~\bibnamefont {Kimura}}, \bibinfo {author} {\bibfnamefont
  {N.}~\bibnamefont {Michel}}, \bibinfo {author} {\bibfnamefont
  {K.}~\bibnamefont {Ogata}}, \bibinfo {author} {\bibfnamefont {C.~X.}\
  \bibnamefont {Yuan}}, \bibinfo {author} {\bibfnamefont {Q.}~\bibnamefont
  {Yuan}}, \bibinfo {author} {\bibfnamefont {G.}~\bibnamefont {Authelet}},
  \bibinfo {author} {\bibfnamefont {H.}~\bibnamefont {Baba}}, \bibinfo {author}
  {\bibfnamefont {C.}~\bibnamefont {Caesar}}, \bibinfo {author} {\bibfnamefont
  {D.}~\bibnamefont {Calvet}}, \bibinfo {author} {\bibfnamefont
  {A.}~\bibnamefont {Delbart}}, \bibinfo {author} {\bibfnamefont
  {M.}~\bibnamefont {Dozono}}, \bibinfo {author} {\bibfnamefont
  {J.}~\bibnamefont {Feng}}, \bibinfo {author} {\bibfnamefont {F.}~\bibnamefont
  {Flavigny}}, \bibinfo {author} {\bibfnamefont {J.-M.}\ \bibnamefont
  {Gheller}}, \bibinfo {author} {\bibfnamefont {J.}~\bibnamefont {Gibelin}},
  \bibinfo {author} {\bibfnamefont {A.}~\bibnamefont {Giganon}}, \bibinfo
  {author} {\bibfnamefont {A.}~\bibnamefont {Gillibert}}, \bibinfo {author}
  {\bibfnamefont {K.}~\bibnamefont {Hasegawa}}, \bibinfo {author}
  {\bibfnamefont {T.}~\bibnamefont {Isobe}}, \bibinfo {author} {\bibfnamefont
  {Y.}~\bibnamefont {Kanaya}}, \bibinfo {author} {\bibfnamefont
  {S.}~\bibnamefont {Kawakami}}, \bibinfo {author} {\bibfnamefont
  {D.}~\bibnamefont {Kim}}, \bibinfo {author} {\bibfnamefont {Y.}~\bibnamefont
  {Kiyokawa}}, \bibinfo {author} {\bibfnamefont {M.}~\bibnamefont {Kobayashi}},
  \bibinfo {author} {\bibfnamefont {N.}~\bibnamefont {Kobayashi}}, \bibinfo
  {author} {\bibfnamefont {T.}~\bibnamefont {Kobayashi}}, \bibinfo {author}
  {\bibfnamefont {Y.}~\bibnamefont {Kondo}}, \bibinfo {author} {\bibfnamefont
  {Z.}~\bibnamefont {Korkulu}}, \bibinfo {author} {\bibfnamefont
  {S.}~\bibnamefont {Koyama}}, \bibinfo {author} {\bibfnamefont
  {V.}~\bibnamefont {Lapoux}}, \bibinfo {author} {\bibfnamefont
  {Y.}~\bibnamefont {Maeda}}, \bibinfo {author} {\bibfnamefont {F.~M.}\
  \bibnamefont {Marqu{\'e}s}}, \bibinfo {author} {\bibfnamefont
  {T.}~\bibnamefont {Motobayashi}}, \bibinfo {author} {\bibfnamefont
  {T.}~\bibnamefont {Miyazaki}}, \bibinfo {author} {\bibfnamefont
  {T.}~\bibnamefont {Nakamura}}, \bibinfo {author} {\bibfnamefont
  {N.}~\bibnamefont {Nakatsuka}}, \bibinfo {author} {\bibfnamefont
  {Y.}~\bibnamefont {Nishio}}, \bibinfo {author} {\bibfnamefont
  {A.}~\bibnamefont {Obertelli}}, \bibinfo {author} {\bibfnamefont
  {A.}~\bibnamefont {Ohkura}}, \bibinfo {author} {\bibfnamefont {N.~A.}\
  \bibnamefont {Orr}}, \bibinfo {author} {\bibfnamefont {S.}~\bibnamefont
  {Ota}}, \bibinfo {author} {\bibfnamefont {H.}~\bibnamefont {Otsu}}, \bibinfo
  {author} {\bibfnamefont {T.}~\bibnamefont {Ozaki}}, \bibinfo {author}
  {\bibfnamefont {V.}~\bibnamefont {Panin}}, \bibinfo {author} {\bibfnamefont
  {S.}~\bibnamefont {Paschalis}}, \bibinfo {author} {\bibfnamefont {E.~C.}\
  \bibnamefont {Pollacco}}, \bibinfo {author} {\bibfnamefont {S.}~\bibnamefont
  {Reichert}}, \bibinfo {author} {\bibfnamefont {J.-Y.}\ \bibnamefont
  {Rouss{\'e}}}, \bibinfo {author} {\bibfnamefont {A.~T.}\ \bibnamefont
  {Saito}}, \bibinfo {author} {\bibfnamefont {S.}~\bibnamefont {Sakaguchi}},
  \bibinfo {author} {\bibfnamefont {M.}~\bibnamefont {Sako}}, \bibinfo {author}
  {\bibfnamefont {C.}~\bibnamefont {Santamaria}}, \bibinfo {author}
  {\bibfnamefont {M.}~\bibnamefont {Sasano}}, \bibinfo {author} {\bibfnamefont
  {H.}~\bibnamefont {Sato}}, \bibinfo {author} {\bibfnamefont {M.}~\bibnamefont
  {Shikata}}, \bibinfo {author} {\bibfnamefont {Y.}~\bibnamefont {Shimizu}},
  \bibinfo {author} {\bibfnamefont {Y.}~\bibnamefont {Shindo}}, \bibinfo
  {author} {\bibfnamefont {L.}~\bibnamefont {Stuhl}}, \bibinfo {author}
  {\bibfnamefont {T.}~\bibnamefont {Sumikama}}, \bibinfo {author}
  {\bibfnamefont {Y.~L.}\ \bibnamefont {Sun}}, \bibinfo {author} {\bibfnamefont
  {M.}~\bibnamefont {Tabata}}, \bibinfo {author} {\bibfnamefont
  {Y.}~\bibnamefont {Togano}}, \bibinfo {author} {\bibfnamefont
  {J.}~\bibnamefont {Tsubota}}, \bibinfo {author} {\bibfnamefont {F.~R.}\
  \bibnamefont {Xu}}, \bibinfo {author} {\bibfnamefont {J.}~\bibnamefont
  {Yasuda}}, \bibinfo {author} {\bibfnamefont {K.}~\bibnamefont {Yoneda}},
  \bibinfo {author} {\bibfnamefont {J.}~\bibnamefont {Zenihiro}}, \bibinfo
  {author} {\bibfnamefont {S.-G.}\ \bibnamefont {Zhou}}, \bibinfo {author}
  {\bibfnamefont {W.}~\bibnamefont {Zuo}},\ and\ \bibinfo {author}
  {\bibfnamefont {T.}~\bibnamefont {Uesaka}},\ }\href
  {https://doi.org/10.1103/physrevlett.126.082501} {\bibfield  {journal}
  {\bibinfo  {journal} {Phys. Rev. Lett.}\ }\textbf {\bibinfo {volume} {126}},\
  \bibinfo {pages} {082501} (\bibinfo {year} {2021})}\BibitemShut {NoStop}%
\bibitem [{\citenamefont {Pougheon}\ \emph {et~al.}(1986)\citenamefont
  {Pougheon}, \citenamefont {{Guillemaud-Mueller}}, \citenamefont {Quiniou},
  \citenamefont {Laurent}, \citenamefont {Anne}, \citenamefont {Bazin},
  \citenamefont {Bernas}, \citenamefont {Guerreau}, \citenamefont {Jacmart},
  \citenamefont {Hoath}, \citenamefont {Mueller},\ and\ \citenamefont
  {D{\'e}traz}}]{Pougheon1986_EPL2-505}%
  \BibitemOpen
  \bibfield  {author} {\bibinfo {author} {\bibfnamefont {F.}~\bibnamefont
  {Pougheon}}, \bibinfo {author} {\bibfnamefont {D.}~\bibnamefont
  {{Guillemaud-Mueller}}}, \bibinfo {author} {\bibfnamefont {E.}~\bibnamefont
  {Quiniou}}, \bibinfo {author} {\bibfnamefont {M.~G.~S.}\ \bibnamefont
  {Laurent}}, \bibinfo {author} {\bibfnamefont {R.}~\bibnamefont {Anne}},
  \bibinfo {author} {\bibfnamefont {D.}~\bibnamefont {Bazin}}, \bibinfo
  {author} {\bibfnamefont {M.}~\bibnamefont {Bernas}}, \bibinfo {author}
  {\bibfnamefont {D.}~\bibnamefont {Guerreau}}, \bibinfo {author}
  {\bibfnamefont {J.~C.}\ \bibnamefont {Jacmart}}, \bibinfo {author}
  {\bibfnamefont {S.~D.}\ \bibnamefont {Hoath}}, \bibinfo {author}
  {\bibfnamefont {A.~C.}\ \bibnamefont {Mueller}},\ and\ \bibinfo {author}
  {\bibfnamefont {C.}~\bibnamefont {D{\'e}traz}},\ }\href
  {https://doi.org/10.1209/0295-5075/2/7/003} {\bibfield  {journal} {\bibinfo
  {journal} {Europhys. Lett.}\ }\textbf {\bibinfo {volume} {2}},\ \bibinfo
  {pages} {505} (\bibinfo {year} {1986})}\BibitemShut {NoStop}%
\bibitem [{\citenamefont {Tanaka}\ \emph {et~al.}(2010)\citenamefont {Tanaka},
  \citenamefont {Yamaguchi}, \citenamefont {Suzuki}, \citenamefont {Ohtsubo},
  \citenamefont {Fukuda}, \citenamefont {Nishimura}, \citenamefont {Takechi},
  \citenamefont {Ogata}, \citenamefont {Ozawa}, \citenamefont {Izumikawa},
  \citenamefont {Aiba}, \citenamefont {Aoi}, \citenamefont {Baba},
  \citenamefont {Hashizume}, \citenamefont {Inafuku}, \citenamefont {Iwasa},
  \citenamefont {Kobayashi}, \citenamefont {Komuro}, \citenamefont {Kondo},
  \citenamefont {Kubo}, \citenamefont {Kurokawa}, \citenamefont {Matsuyama},
  \citenamefont {Michimasa}, \citenamefont {Motobayashi}, \citenamefont
  {Nakabayashi}, \citenamefont {Nakajima}, \citenamefont {Nakamura},
  \citenamefont {Sakurai}, \citenamefont {Shinoda}, \citenamefont {Shinohara},
  \citenamefont {Suzuki}, \citenamefont {Takeshita}, \citenamefont {Takeuchi},
  \citenamefont {Togano}, \citenamefont {Yamada}, \citenamefont {Yasuno},\ and\
  \citenamefont {Yoshitake}}]{Tanaka2010_PRL104-062701}%
  \BibitemOpen
  \bibfield  {author} {\bibinfo {author} {\bibfnamefont {K.}~\bibnamefont
  {Tanaka}}, \bibinfo {author} {\bibfnamefont {T.}~\bibnamefont {Yamaguchi}},
  \bibinfo {author} {\bibfnamefont {T.}~\bibnamefont {Suzuki}}, \bibinfo
  {author} {\bibfnamefont {T.}~\bibnamefont {Ohtsubo}}, \bibinfo {author}
  {\bibfnamefont {M.}~\bibnamefont {Fukuda}}, \bibinfo {author} {\bibfnamefont
  {D.}~\bibnamefont {Nishimura}}, \bibinfo {author} {\bibfnamefont
  {M.}~\bibnamefont {Takechi}}, \bibinfo {author} {\bibfnamefont
  {K.}~\bibnamefont {Ogata}}, \bibinfo {author} {\bibfnamefont
  {A.}~\bibnamefont {Ozawa}}, \bibinfo {author} {\bibfnamefont
  {T.}~\bibnamefont {Izumikawa}}, \bibinfo {author} {\bibfnamefont
  {T.}~\bibnamefont {Aiba}}, \bibinfo {author} {\bibfnamefont {N.}~\bibnamefont
  {Aoi}}, \bibinfo {author} {\bibfnamefont {H.}~\bibnamefont {Baba}}, \bibinfo
  {author} {\bibfnamefont {Y.}~\bibnamefont {Hashizume}}, \bibinfo {author}
  {\bibfnamefont {K.}~\bibnamefont {Inafuku}}, \bibinfo {author} {\bibfnamefont
  {N.}~\bibnamefont {Iwasa}}, \bibinfo {author} {\bibfnamefont
  {K.}~\bibnamefont {Kobayashi}}, \bibinfo {author} {\bibfnamefont
  {M.}~\bibnamefont {Komuro}}, \bibinfo {author} {\bibfnamefont
  {Y.}~\bibnamefont {Kondo}}, \bibinfo {author} {\bibfnamefont
  {T.}~\bibnamefont {Kubo}}, \bibinfo {author} {\bibfnamefont {M.}~\bibnamefont
  {Kurokawa}}, \bibinfo {author} {\bibfnamefont {T.}~\bibnamefont {Matsuyama}},
  \bibinfo {author} {\bibfnamefont {S.}~\bibnamefont {Michimasa}}, \bibinfo
  {author} {\bibfnamefont {T.}~\bibnamefont {Motobayashi}}, \bibinfo {author}
  {\bibfnamefont {T.}~\bibnamefont {Nakabayashi}}, \bibinfo {author}
  {\bibfnamefont {S.}~\bibnamefont {Nakajima}}, \bibinfo {author}
  {\bibfnamefont {T.}~\bibnamefont {Nakamura}}, \bibinfo {author}
  {\bibfnamefont {H.}~\bibnamefont {Sakurai}}, \bibinfo {author} {\bibfnamefont
  {R.}~\bibnamefont {Shinoda}}, \bibinfo {author} {\bibfnamefont
  {M.}~\bibnamefont {Shinohara}}, \bibinfo {author} {\bibfnamefont
  {H.}~\bibnamefont {Suzuki}}, \bibinfo {author} {\bibfnamefont
  {E.}~\bibnamefont {Takeshita}}, \bibinfo {author} {\bibfnamefont
  {S.}~\bibnamefont {Takeuchi}}, \bibinfo {author} {\bibfnamefont
  {Y.}~\bibnamefont {Togano}}, \bibinfo {author} {\bibfnamefont
  {K.}~\bibnamefont {Yamada}}, \bibinfo {author} {\bibfnamefont
  {T.}~\bibnamefont {Yasuno}},\ and\ \bibinfo {author} {\bibfnamefont
  {M.}~\bibnamefont {Yoshitake}},\ }\href
  {https://doi.org/10.1103/PhysRevLett.104.062701} {\bibfield  {journal}
  {\bibinfo  {journal} {Phys. Rev. Lett.}\ }\textbf {\bibinfo {volume} {104}},\
  \bibinfo {pages} {062701} (\bibinfo {year} {2010})}\BibitemShut {NoStop}%
\bibitem [{\citenamefont {Sun}\ \emph {et~al.}(2018)\citenamefont {Sun},
  \citenamefont {Zhao},\ and\ \citenamefont {Zhou}}]{Sun2018_PLB785-530}%
  \BibitemOpen
  \bibfield  {author} {\bibinfo {author} {\bibfnamefont {X.-X.}\ \bibnamefont
  {Sun}}, \bibinfo {author} {\bibfnamefont {J.}~\bibnamefont {Zhao}},\ and\
  \bibinfo {author} {\bibfnamefont {S.-G.}\ \bibnamefont {Zhou}},\ }\href
  {https://doi.org/10.1016/j.physletb.2018.08.071} {\bibfield  {journal}
  {\bibinfo  {journal} {Phys. Lett. B}\ }\textbf {\bibinfo {volume} {785}},\
  \bibinfo {pages} {530} (\bibinfo {year} {2018})},\ \Eprint
  {https://arxiv.org/abs/1807.04991} {arXiv:1807.04991 [nucl-th]} \BibitemShut
  {NoStop}%
\bibitem [{\citenamefont {Stanoiu}\ \emph {et~al.}(2008)\citenamefont
  {Stanoiu}, \citenamefont {Sohler}, \citenamefont {Sorlin}, \citenamefont
  {Azaiez}, \citenamefont {Dombr{\'a}di}, \citenamefont {Brown}, \citenamefont
  {Belleguic}, \citenamefont {Borcea}, \citenamefont {Bourgeois}, \citenamefont
  {Dlouhy}, \citenamefont {Elekes}, \citenamefont {F{\"u}l{\"o}p},
  \citenamefont {Gr{\'e}vy}, \citenamefont {{Guillemaud-Mueller}},
  \citenamefont {Ibrahim}, \citenamefont {Kerek}, \citenamefont
  {Krasznahorkay}, \citenamefont {Lewitowicz}, \citenamefont {Lukyanov},
  \citenamefont {Mandal}, \citenamefont {Mr{\'a}zek}, \citenamefont {Negoita},
  \citenamefont {Penionzhkevich}, \citenamefont {Podoly{\'a}k}, \citenamefont
  {{Roussel-Chomaz}}, \citenamefont {{Saint-Laurent}}, \citenamefont
  {Savajols}, \citenamefont {Sletten}, \citenamefont {Tim{\'a}r}, \citenamefont
  {Timis},\ and\ \citenamefont {Yamamoto}}]{Stanoiu2008_PRC78-034315}%
  \BibitemOpen
  \bibfield  {author} {\bibinfo {author} {\bibfnamefont {M.}~\bibnamefont
  {Stanoiu}}, \bibinfo {author} {\bibfnamefont {D.}~\bibnamefont {Sohler}},
  \bibinfo {author} {\bibfnamefont {O.}~\bibnamefont {Sorlin}}, \bibinfo
  {author} {\bibfnamefont {F.}~\bibnamefont {Azaiez}}, \bibinfo {author}
  {\bibfnamefont {{\relax Zs}.}~\bibnamefont {Dombr{\'a}di}}, \bibinfo {author}
  {\bibfnamefont {B.~A.}\ \bibnamefont {Brown}}, \bibinfo {author}
  {\bibfnamefont {M.}~\bibnamefont {Belleguic}}, \bibinfo {author}
  {\bibfnamefont {C.}~\bibnamefont {Borcea}}, \bibinfo {author} {\bibfnamefont
  {C.}~\bibnamefont {Bourgeois}}, \bibinfo {author} {\bibfnamefont
  {Z.}~\bibnamefont {Dlouhy}}, \bibinfo {author} {\bibfnamefont
  {Z.}~\bibnamefont {Elekes}}, \bibinfo {author} {\bibfnamefont {{\relax
  Zs}.}~\bibnamefont {F{\"u}l{\"o}p}}, \bibinfo {author} {\bibfnamefont
  {S.}~\bibnamefont {Gr{\'e}vy}}, \bibinfo {author} {\bibfnamefont
  {D.}~\bibnamefont {{Guillemaud-Mueller}}}, \bibinfo {author} {\bibfnamefont
  {F.}~\bibnamefont {Ibrahim}}, \bibinfo {author} {\bibfnamefont
  {A.}~\bibnamefont {Kerek}}, \bibinfo {author} {\bibfnamefont
  {A.}~\bibnamefont {Krasznahorkay}}, \bibinfo {author} {\bibfnamefont
  {M.}~\bibnamefont {Lewitowicz}}, \bibinfo {author} {\bibfnamefont {S.~M.}\
  \bibnamefont {Lukyanov}}, \bibinfo {author} {\bibfnamefont {S.}~\bibnamefont
  {Mandal}}, \bibinfo {author} {\bibfnamefont {J.}~\bibnamefont {Mr{\'a}zek}},
  \bibinfo {author} {\bibfnamefont {F.}~\bibnamefont {Negoita}}, \bibinfo
  {author} {\bibfnamefont {Y.-E.}\ \bibnamefont {Penionzhkevich}}, \bibinfo
  {author} {\bibfnamefont {{\relax Zs}.}~\bibnamefont {Podoly{\'a}k}}, \bibinfo
  {author} {\bibfnamefont {P.}~\bibnamefont {{Roussel-Chomaz}}}, \bibinfo
  {author} {\bibfnamefont {M.~G.}\ \bibnamefont {{Saint-Laurent}}}, \bibinfo
  {author} {\bibfnamefont {H.}~\bibnamefont {Savajols}}, \bibinfo {author}
  {\bibfnamefont {G.}~\bibnamefont {Sletten}}, \bibinfo {author} {\bibfnamefont
  {J.}~\bibnamefont {Tim{\'a}r}}, \bibinfo {author} {\bibfnamefont
  {C.}~\bibnamefont {Timis}},\ and\ \bibinfo {author} {\bibfnamefont
  {A.}~\bibnamefont {Yamamoto}},\ }\href
  {https://doi.org/10.1103/PhysRevC.78.034315} {\bibfield  {journal} {\bibinfo
  {journal} {Phys. Rev. C}\ }\textbf {\bibinfo {volume} {78}},\ \bibinfo
  {pages} {034315} (\bibinfo {year} {2008})}\BibitemShut {NoStop}%
\bibitem [{\citenamefont {Elekes}\ \emph {et~al.}(2004)\citenamefont {Elekes},
  \citenamefont {Dombr{\'a}di}, \citenamefont {Krasznahorkay}, \citenamefont
  {Baba}, \citenamefont {Csatl{\'o}s}, \citenamefont {Csige}, \citenamefont
  {Fukuda}, \citenamefont {F{\"u}l{\"o}p}, \citenamefont {G{\'a}csi},
  \citenamefont {Guly{\'a}s}, \citenamefont {Iwasa}, \citenamefont {Kinugawa},
  \citenamefont {Kubono}, \citenamefont {Kurokawa}, \citenamefont {Liu},
  \citenamefont {Michimasa}, \citenamefont {Minemura}, \citenamefont
  {Motobayashi}, \citenamefont {Ozawa}, \citenamefont {Saito}, \citenamefont
  {Shimoura}, \citenamefont {Takeuchi}, \citenamefont {Tanihata}, \citenamefont
  {Thirolf}, \citenamefont {Yanagisawa},\ and\ \citenamefont
  {Yoshida}}]{Elekes2004_PLB586-34}%
  \BibitemOpen
  \bibfield  {author} {\bibinfo {author} {\bibfnamefont {Z.}~\bibnamefont
  {Elekes}}, \bibinfo {author} {\bibfnamefont {{\relax Zs}.}~\bibnamefont
  {Dombr{\'a}di}}, \bibinfo {author} {\bibfnamefont {A.}~\bibnamefont
  {Krasznahorkay}}, \bibinfo {author} {\bibfnamefont {H.}~\bibnamefont {Baba}},
  \bibinfo {author} {\bibfnamefont {M.}~\bibnamefont {Csatl{\'o}s}}, \bibinfo
  {author} {\bibfnamefont {L.}~\bibnamefont {Csige}}, \bibinfo {author}
  {\bibfnamefont {N.}~\bibnamefont {Fukuda}}, \bibinfo {author} {\bibfnamefont
  {{\relax Zs}.}~\bibnamefont {F{\"u}l{\"o}p}}, \bibinfo {author}
  {\bibfnamefont {Z.}~\bibnamefont {G{\'a}csi}}, \bibinfo {author}
  {\bibfnamefont {J.}~\bibnamefont {Guly{\'a}s}}, \bibinfo {author}
  {\bibfnamefont {N.}~\bibnamefont {Iwasa}}, \bibinfo {author} {\bibfnamefont
  {H.}~\bibnamefont {Kinugawa}}, \bibinfo {author} {\bibfnamefont
  {S.}~\bibnamefont {Kubono}}, \bibinfo {author} {\bibfnamefont
  {M.}~\bibnamefont {Kurokawa}}, \bibinfo {author} {\bibfnamefont
  {X.}~\bibnamefont {Liu}}, \bibinfo {author} {\bibfnamefont {S.}~\bibnamefont
  {Michimasa}}, \bibinfo {author} {\bibfnamefont {T.}~\bibnamefont {Minemura}},
  \bibinfo {author} {\bibfnamefont {T.}~\bibnamefont {Motobayashi}}, \bibinfo
  {author} {\bibfnamefont {A.}~\bibnamefont {Ozawa}}, \bibinfo {author}
  {\bibfnamefont {A.}~\bibnamefont {Saito}}, \bibinfo {author} {\bibfnamefont
  {S.}~\bibnamefont {Shimoura}}, \bibinfo {author} {\bibfnamefont
  {S.}~\bibnamefont {Takeuchi}}, \bibinfo {author} {\bibfnamefont
  {I.}~\bibnamefont {Tanihata}}, \bibinfo {author} {\bibfnamefont
  {P.}~\bibnamefont {Thirolf}}, \bibinfo {author} {\bibfnamefont
  {Y.}~\bibnamefont {Yanagisawa}},\ and\ \bibinfo {author} {\bibfnamefont
  {K.}~\bibnamefont {Yoshida}},\ }\href
  {https://doi.org/10.1016/j.physletb.2004.02.022} {\bibfield  {journal}
  {\bibinfo  {journal} {Phys. Lett. B}\ }\textbf {\bibinfo {volume} {586}},\
  \bibinfo {pages} {34} (\bibinfo {year} {2004})}\BibitemShut {NoStop}%
\bibitem [{\citenamefont {Petri}\ \emph {et~al.}(2011)\citenamefont {Petri},
  \citenamefont {Fallon}, \citenamefont {Macchiavelli}, \citenamefont
  {Paschalis}, \citenamefont {Starosta}, \citenamefont {Baugher}, \citenamefont
  {Bazin}, \citenamefont {Cartegni}, \citenamefont {Clark}, \citenamefont
  {Crawford}, \citenamefont {Cromaz}, \citenamefont {Dewald}, \citenamefont
  {Gade}, \citenamefont {Grinyer}, \citenamefont {Gros}, \citenamefont
  {Hackstein}, \citenamefont {Jeppesen}, \citenamefont {Lee}, \citenamefont
  {McDaniel}, \citenamefont {Miller}, \citenamefont {Rajabali}, \citenamefont
  {Ratkiewicz}, \citenamefont {Rother}, \citenamefont {Voss}, \citenamefont
  {Walsh}, \citenamefont {Weisshaar}, \citenamefont {Wiedeking},\ and\
  \citenamefont {Brown}}]{Petri2011_PRL107-102501}%
  \BibitemOpen
  \bibfield  {author} {\bibinfo {author} {\bibfnamefont {M.}~\bibnamefont
  {Petri}}, \bibinfo {author} {\bibfnamefont {P.}~\bibnamefont {Fallon}},
  \bibinfo {author} {\bibfnamefont {A.~O.}\ \bibnamefont {Macchiavelli}},
  \bibinfo {author} {\bibfnamefont {S.}~\bibnamefont {Paschalis}}, \bibinfo
  {author} {\bibfnamefont {K.}~\bibnamefont {Starosta}}, \bibinfo {author}
  {\bibfnamefont {T.}~\bibnamefont {Baugher}}, \bibinfo {author} {\bibfnamefont
  {D.}~\bibnamefont {Bazin}}, \bibinfo {author} {\bibfnamefont
  {L.}~\bibnamefont {Cartegni}}, \bibinfo {author} {\bibfnamefont {R.~M.}\
  \bibnamefont {Clark}}, \bibinfo {author} {\bibfnamefont {H.~L.}\ \bibnamefont
  {Crawford}}, \bibinfo {author} {\bibfnamefont {M.}~\bibnamefont {Cromaz}},
  \bibinfo {author} {\bibfnamefont {A.}~\bibnamefont {Dewald}}, \bibinfo
  {author} {\bibfnamefont {A.}~\bibnamefont {Gade}}, \bibinfo {author}
  {\bibfnamefont {G.~F.}\ \bibnamefont {Grinyer}}, \bibinfo {author}
  {\bibfnamefont {S.}~\bibnamefont {Gros}}, \bibinfo {author} {\bibfnamefont
  {M.}~\bibnamefont {Hackstein}}, \bibinfo {author} {\bibfnamefont {H.~B.}\
  \bibnamefont {Jeppesen}}, \bibinfo {author} {\bibfnamefont {I.~Y.}\
  \bibnamefont {Lee}}, \bibinfo {author} {\bibfnamefont {S.}~\bibnamefont
  {McDaniel}}, \bibinfo {author} {\bibfnamefont {D.}~\bibnamefont {Miller}},
  \bibinfo {author} {\bibfnamefont {M.~M.}\ \bibnamefont {Rajabali}}, \bibinfo
  {author} {\bibfnamefont {A.}~\bibnamefont {Ratkiewicz}}, \bibinfo {author}
  {\bibfnamefont {W.}~\bibnamefont {Rother}}, \bibinfo {author} {\bibfnamefont
  {P.}~\bibnamefont {Voss}}, \bibinfo {author} {\bibfnamefont {K.~A.}\
  \bibnamefont {Walsh}}, \bibinfo {author} {\bibfnamefont {D.}~\bibnamefont
  {Weisshaar}}, \bibinfo {author} {\bibfnamefont {M.}~\bibnamefont
  {Wiedeking}},\ and\ \bibinfo {author} {\bibfnamefont {B.~A.}\ \bibnamefont
  {Brown}},\ }\href {https://doi.org/10.1103/PhysRevLett.107.102501} {\bibfield
   {journal} {\bibinfo  {journal} {Phys. Rev. Lett.}\ }\textbf {\bibinfo
  {volume} {107}},\ \bibinfo {pages} {102501} (\bibinfo {year}
  {2011})}\BibitemShut {NoStop}%
\bibitem [{\citenamefont {Ozawa}\ \emph {et~al.}(2000)\citenamefont {Ozawa},
  \citenamefont {Kobayashi}, \citenamefont {Suzuki}, \citenamefont {Yoshida},\
  and\ \citenamefont {Tanihata}}]{Ozawa2000_PRL84-5493}%
  \BibitemOpen
  \bibfield  {author} {\bibinfo {author} {\bibfnamefont {A.}~\bibnamefont
  {Ozawa}}, \bibinfo {author} {\bibfnamefont {T.}~\bibnamefont {Kobayashi}},
  \bibinfo {author} {\bibfnamefont {T.}~\bibnamefont {Suzuki}}, \bibinfo
  {author} {\bibfnamefont {K.}~\bibnamefont {Yoshida}},\ and\ \bibinfo {author}
  {\bibfnamefont {I.}~\bibnamefont {Tanihata}},\ }\href
  {https://doi.org/10.1103/PhysRevLett.84.5493} {\bibfield  {journal} {\bibinfo
   {journal} {Phys. Rev. Lett.}\ }\textbf {\bibinfo {volume} {84}},\ \bibinfo
  {pages} {5493} (\bibinfo {year} {2000})}\BibitemShut {NoStop}%
\bibitem [{\citenamefont {Kanungo}\ \emph {et~al.}(2009)\citenamefont
  {Kanungo}, \citenamefont {Nociforo}, \citenamefont {Prochazka}, \citenamefont
  {Aumann}, \citenamefont {Boutin}, \citenamefont {{Cortina-Gil}},
  \citenamefont {Davids}, \citenamefont {Diakaki}, \citenamefont {Farinon},
  \citenamefont {Geissel}, \citenamefont {Gernh{\"a}user}, \citenamefont
  {Gerl}, \citenamefont {Janik}, \citenamefont {Jonson}, \citenamefont
  {Kindler}, \citenamefont {Kn{\"o}bel}, \citenamefont {Kr{\"u}cken},
  \citenamefont {Lantz}, \citenamefont {Lenske}, \citenamefont {Litvinov},
  \citenamefont {Lommel}, \citenamefont {Mahata}, \citenamefont {Maierbeck},
  \citenamefont {Musumarra}, \citenamefont {Nilsson}, \citenamefont {Otsuka},
  \citenamefont {Perro}, \citenamefont {Scheidenberger}, \citenamefont {Sitar},
  \citenamefont {Strmen}, \citenamefont {Sun}, \citenamefont {Szarka},
  \citenamefont {Tanihata}, \citenamefont {Utsuno}, \citenamefont {Weick},\
  and\ \citenamefont {Winkler}}]{Kanungo2009_PRL102-152501}%
  \BibitemOpen
  \bibfield  {author} {\bibinfo {author} {\bibfnamefont {R.}~\bibnamefont
  {Kanungo}}, \bibinfo {author} {\bibfnamefont {C.}~\bibnamefont {Nociforo}},
  \bibinfo {author} {\bibfnamefont {A.}~\bibnamefont {Prochazka}}, \bibinfo
  {author} {\bibfnamefont {T.}~\bibnamefont {Aumann}}, \bibinfo {author}
  {\bibfnamefont {D.}~\bibnamefont {Boutin}}, \bibinfo {author} {\bibfnamefont
  {D.}~\bibnamefont {{Cortina-Gil}}}, \bibinfo {author} {\bibfnamefont
  {B.}~\bibnamefont {Davids}}, \bibinfo {author} {\bibfnamefont
  {M.}~\bibnamefont {Diakaki}}, \bibinfo {author} {\bibfnamefont
  {F.}~\bibnamefont {Farinon}}, \bibinfo {author} {\bibfnamefont
  {H.}~\bibnamefont {Geissel}}, \bibinfo {author} {\bibfnamefont
  {R.}~\bibnamefont {Gernh{\"a}user}}, \bibinfo {author} {\bibfnamefont
  {J.}~\bibnamefont {Gerl}}, \bibinfo {author} {\bibfnamefont {R.}~\bibnamefont
  {Janik}}, \bibinfo {author} {\bibfnamefont {B.}~\bibnamefont {Jonson}},
  \bibinfo {author} {\bibfnamefont {B.}~\bibnamefont {Kindler}}, \bibinfo
  {author} {\bibfnamefont {R.}~\bibnamefont {Kn{\"o}bel}}, \bibinfo {author}
  {\bibfnamefont {R.}~\bibnamefont {Kr{\"u}cken}}, \bibinfo {author}
  {\bibfnamefont {M.}~\bibnamefont {Lantz}}, \bibinfo {author} {\bibfnamefont
  {H.}~\bibnamefont {Lenske}}, \bibinfo {author} {\bibfnamefont
  {Y.}~\bibnamefont {Litvinov}}, \bibinfo {author} {\bibfnamefont
  {B.}~\bibnamefont {Lommel}}, \bibinfo {author} {\bibfnamefont
  {K.}~\bibnamefont {Mahata}}, \bibinfo {author} {\bibfnamefont
  {P.}~\bibnamefont {Maierbeck}}, \bibinfo {author} {\bibfnamefont
  {A.}~\bibnamefont {Musumarra}}, \bibinfo {author} {\bibfnamefont
  {T.}~\bibnamefont {Nilsson}}, \bibinfo {author} {\bibfnamefont
  {T.}~\bibnamefont {Otsuka}}, \bibinfo {author} {\bibfnamefont
  {C.}~\bibnamefont {Perro}}, \bibinfo {author} {\bibfnamefont
  {C.}~\bibnamefont {Scheidenberger}}, \bibinfo {author} {\bibfnamefont
  {B.}~\bibnamefont {Sitar}}, \bibinfo {author} {\bibfnamefont
  {P.}~\bibnamefont {Strmen}}, \bibinfo {author} {\bibfnamefont
  {B.}~\bibnamefont {Sun}}, \bibinfo {author} {\bibfnamefont {I.}~\bibnamefont
  {Szarka}}, \bibinfo {author} {\bibfnamefont {I.}~\bibnamefont {Tanihata}},
  \bibinfo {author} {\bibfnamefont {Y.}~\bibnamefont {Utsuno}}, \bibinfo
  {author} {\bibfnamefont {H.}~\bibnamefont {Weick}},\ and\ \bibinfo {author}
  {\bibfnamefont {M.}~\bibnamefont {Winkler}},\ }\href
  {https://doi.org/10.1103/PhysRevLett.102.152501} {\bibfield  {journal}
  {\bibinfo  {journal} {Phys. Rev. Lett.}\ }\textbf {\bibinfo {volume} {102}},\
  \bibinfo {pages} {152501} (\bibinfo {year} {2009})}\BibitemShut {NoStop}%
\bibitem [{\citenamefont {Tshoo}\ \emph {et~al.}(2012)\citenamefont {Tshoo},
  \citenamefont {Satou}, \citenamefont {Bhang} \emph
  {et~al.}}]{Tshoo2012_PRL109-22501}%
  \BibitemOpen
  \bibfield  {author} {\bibinfo {author} {\bibfnamefont {K.}~\bibnamefont
  {Tshoo}}, \bibinfo {author} {\bibfnamefont {Y.}~\bibnamefont {Satou}},
  \bibinfo {author} {\bibfnamefont {H.}~\bibnamefont {Bhang}}, \emph {et~al.},\
  }\href {https://doi.org/10.1103/PhysRevLett.109.022501} {\bibfield  {journal}
  {\bibinfo  {journal} {Phys. Rev. Lett.}\ }\textbf {\bibinfo {volume} {109}},\
  \bibinfo {pages} {22501} (\bibinfo {year} {2012})},\ \Eprint
  {https://arxiv.org/abs/1205.5657} {arXiv:1205.5657 [nucl-ex]} \BibitemShut
  {NoStop}%
\bibitem [{\citenamefont {Togano}\ \emph {et~al.}(2016)\citenamefont {Togano},
  \citenamefont {Nakamura}, \citenamefont {Kondo}, \citenamefont {Tostevin},
  \citenamefont {Saito}, \citenamefont {Gibelin}, \citenamefont {Orr},
  \citenamefont {Achouri}, \citenamefont {Aumann}, \citenamefont {Baba},
  \citenamefont {Delaunay}, \citenamefont {Doornenbal}, \citenamefont {Fukuda},
  \citenamefont {Hwang}, \citenamefont {Inabe}, \citenamefont {Isobe},
  \citenamefont {Kameda}, \citenamefont {Kanno}, \citenamefont {Kim},
  \citenamefont {Kobayashi}, \citenamefont {Kobayashi}, \citenamefont {Kubo},
  \citenamefont {Leblond}, \citenamefont {Lee}, \citenamefont {Marques},
  \citenamefont {Minakata}, \citenamefont {Motobayashi}, \citenamefont {Murai},
  \citenamefont {Murakami}, \citenamefont {Muto}, \citenamefont {Nakashima},
  \citenamefont {Nakatsuka}, \citenamefont {Navin}, \citenamefont {Nishi},
  \citenamefont {Ogoshi}, \citenamefont {Otsu}, \citenamefont {Sato},
  \citenamefont {Satou}, \citenamefont {Shimizu}, \citenamefont {Suzuki},
  \citenamefont {Takahashi}, \citenamefont {Takeda}, \citenamefont {Takeuchi},
  \citenamefont {Tanaka}, \citenamefont {Tuff}, \citenamefont {Vandebrouck},\
  and\ \citenamefont {Yoneda}}]{Togano2016_PLB761-412}%
  \BibitemOpen
  \bibfield  {author} {\bibinfo {author} {\bibfnamefont {Y.}~\bibnamefont
  {Togano}}, \bibinfo {author} {\bibfnamefont {T.}~\bibnamefont {Nakamura}},
  \bibinfo {author} {\bibfnamefont {Y.}~\bibnamefont {Kondo}}, \bibinfo
  {author} {\bibfnamefont {J.}~\bibnamefont {Tostevin}}, \bibinfo {author}
  {\bibfnamefont {A.}~\bibnamefont {Saito}}, \bibinfo {author} {\bibfnamefont
  {J.}~\bibnamefont {Gibelin}}, \bibinfo {author} {\bibfnamefont
  {N.}~\bibnamefont {Orr}}, \bibinfo {author} {\bibfnamefont {N.}~\bibnamefont
  {Achouri}}, \bibinfo {author} {\bibfnamefont {T.}~\bibnamefont {Aumann}},
  \bibinfo {author} {\bibfnamefont {H.}~\bibnamefont {Baba}}, \bibinfo {author}
  {\bibfnamefont {F.}~\bibnamefont {Delaunay}}, \bibinfo {author}
  {\bibfnamefont {P.}~\bibnamefont {Doornenbal}}, \bibinfo {author}
  {\bibfnamefont {N.}~\bibnamefont {Fukuda}}, \bibinfo {author} {\bibfnamefont
  {J.}~\bibnamefont {Hwang}}, \bibinfo {author} {\bibfnamefont
  {N.}~\bibnamefont {Inabe}}, \bibinfo {author} {\bibfnamefont
  {T.}~\bibnamefont {Isobe}}, \bibinfo {author} {\bibfnamefont
  {D.}~\bibnamefont {Kameda}}, \bibinfo {author} {\bibfnamefont
  {D.}~\bibnamefont {Kanno}}, \bibinfo {author} {\bibfnamefont
  {S.}~\bibnamefont {Kim}}, \bibinfo {author} {\bibfnamefont {N.}~\bibnamefont
  {Kobayashi}}, \bibinfo {author} {\bibfnamefont {T.}~\bibnamefont
  {Kobayashi}}, \bibinfo {author} {\bibfnamefont {T.}~\bibnamefont {Kubo}},
  \bibinfo {author} {\bibfnamefont {S.}~\bibnamefont {Leblond}}, \bibinfo
  {author} {\bibfnamefont {J.}~\bibnamefont {Lee}}, \bibinfo {author}
  {\bibfnamefont {F.}~\bibnamefont {Marques}}, \bibinfo {author} {\bibfnamefont
  {R.}~\bibnamefont {Minakata}}, \bibinfo {author} {\bibfnamefont
  {T.}~\bibnamefont {Motobayashi}}, \bibinfo {author} {\bibfnamefont
  {D.}~\bibnamefont {Murai}}, \bibinfo {author} {\bibfnamefont
  {T.}~\bibnamefont {Murakami}}, \bibinfo {author} {\bibfnamefont
  {K.}~\bibnamefont {Muto}}, \bibinfo {author} {\bibfnamefont {T.}~\bibnamefont
  {Nakashima}}, \bibinfo {author} {\bibfnamefont {N.}~\bibnamefont
  {Nakatsuka}}, \bibinfo {author} {\bibfnamefont {A.}~\bibnamefont {Navin}},
  \bibinfo {author} {\bibfnamefont {S.}~\bibnamefont {Nishi}}, \bibinfo
  {author} {\bibfnamefont {S.}~\bibnamefont {Ogoshi}}, \bibinfo {author}
  {\bibfnamefont {H.}~\bibnamefont {Otsu}}, \bibinfo {author} {\bibfnamefont
  {H.}~\bibnamefont {Sato}}, \bibinfo {author} {\bibfnamefont {Y.}~\bibnamefont
  {Satou}}, \bibinfo {author} {\bibfnamefont {Y.}~\bibnamefont {Shimizu}},
  \bibinfo {author} {\bibfnamefont {H.}~\bibnamefont {Suzuki}}, \bibinfo
  {author} {\bibfnamefont {K.}~\bibnamefont {Takahashi}}, \bibinfo {author}
  {\bibfnamefont {H.}~\bibnamefont {Takeda}}, \bibinfo {author} {\bibfnamefont
  {S.}~\bibnamefont {Takeuchi}}, \bibinfo {author} {\bibfnamefont
  {R.}~\bibnamefont {Tanaka}}, \bibinfo {author} {\bibfnamefont
  {A.}~\bibnamefont {Tuff}}, \bibinfo {author} {\bibfnamefont {M.}~\bibnamefont
  {Vandebrouck}},\ and\ \bibinfo {author} {\bibfnamefont {K.}~\bibnamefont
  {Yoneda}},\ }\href {https://doi.org/10.1016/j.physletb.2016.08.062}
  {\bibfield  {journal} {\bibinfo  {journal} {Phys. Lett. B}\ }\textbf
  {\bibinfo {volume} {761}},\ \bibinfo {pages} {412} (\bibinfo {year}
  {2016})}\BibitemShut {NoStop}%
\bibitem [{\citenamefont {Nagahisa}\ and\ \citenamefont
  {Horiuchi}(2018)}]{Nagahisa2018_PRC97-054614}%
  \BibitemOpen
  \bibfield  {author} {\bibinfo {author} {\bibfnamefont {T.}~\bibnamefont
  {Nagahisa}}\ and\ \bibinfo {author} {\bibfnamefont {W.}~\bibnamefont
  {Horiuchi}},\ }\href {https://doi.org/10.1103/PhysRevC.97.054614} {\bibfield
  {journal} {\bibinfo  {journal} {Phys. Rev. C}\ }\textbf {\bibinfo {volume}
  {97}},\ \bibinfo {pages} {054614} (\bibinfo {year} {2018})}\BibitemShut
  {NoStop}%
\bibitem [{\citenamefont {Gaudefroy}\ \emph {et~al.}(2012)\citenamefont
  {Gaudefroy}, \citenamefont {Mittig}, \citenamefont {Orr}, \citenamefont
  {Varet}, \citenamefont {Chartier}, \citenamefont {{Roussel-Chomaz}},
  \citenamefont {Ebran}, \citenamefont {{Fern{\'a}ndez-Dom{\'i}nguez}},
  \citenamefont {Fr{\'e}mont}, \citenamefont {Gangnant}, \citenamefont
  {Gillibert}, \citenamefont {Gr{\'e}vy}, \citenamefont {Libin}, \citenamefont
  {Maslov}, \citenamefont {Paschalis}, \citenamefont {Pietras}, \citenamefont
  {Penionzhkevich}, \citenamefont {Spitaels},\ and\ \citenamefont
  {Villari}}]{Gaudefroy2012_PRL109-202503}%
  \BibitemOpen
  \bibfield  {author} {\bibinfo {author} {\bibfnamefont {L.}~\bibnamefont
  {Gaudefroy}}, \bibinfo {author} {\bibfnamefont {W.}~\bibnamefont {Mittig}},
  \bibinfo {author} {\bibfnamefont {N.~A.}\ \bibnamefont {Orr}}, \bibinfo
  {author} {\bibfnamefont {S.}~\bibnamefont {Varet}}, \bibinfo {author}
  {\bibfnamefont {M.}~\bibnamefont {Chartier}}, \bibinfo {author}
  {\bibfnamefont {P.}~\bibnamefont {{Roussel-Chomaz}}}, \bibinfo {author}
  {\bibfnamefont {J.~P.}\ \bibnamefont {Ebran}}, \bibinfo {author}
  {\bibfnamefont {B.}~\bibnamefont {{Fern{\'a}ndez-Dom{\'i}nguez}}}, \bibinfo
  {author} {\bibfnamefont {G.}~\bibnamefont {Fr{\'e}mont}}, \bibinfo {author}
  {\bibfnamefont {P.}~\bibnamefont {Gangnant}}, \bibinfo {author}
  {\bibfnamefont {A.}~\bibnamefont {Gillibert}}, \bibinfo {author}
  {\bibfnamefont {S.}~\bibnamefont {Gr{\'e}vy}}, \bibinfo {author}
  {\bibfnamefont {J.~F.}\ \bibnamefont {Libin}}, \bibinfo {author}
  {\bibfnamefont {V.~A.}\ \bibnamefont {Maslov}}, \bibinfo {author}
  {\bibfnamefont {S.}~\bibnamefont {Paschalis}}, \bibinfo {author}
  {\bibfnamefont {B.}~\bibnamefont {Pietras}}, \bibinfo {author} {\bibfnamefont
  {Y.-E.}\ \bibnamefont {Penionzhkevich}}, \bibinfo {author} {\bibfnamefont
  {C.}~\bibnamefont {Spitaels}},\ and\ \bibinfo {author} {\bibfnamefont
  {A.~C.~C.}\ \bibnamefont {Villari}},\ }\href
  {https://doi.org/10.1103/PhysRevLett.109.202503} {\bibfield  {journal}
  {\bibinfo  {journal} {Phys. Rev. Lett.}\ }\textbf {\bibinfo {volume} {109}},\
  \bibinfo {pages} {202503} (\bibinfo {year} {2012})}\BibitemShut {NoStop}%
\bibitem [{\citenamefont {Wang}\ \emph {et~al.}(2021)\citenamefont {Wang},
  \citenamefont {Huang}, \citenamefont {Kondev}, \citenamefont {Audi},\ and\
  \citenamefont {Naimi}}]{Wang2021_ChinPhysC45-030003}%
  \BibitemOpen
  \bibfield  {author} {\bibinfo {author} {\bibfnamefont {M.}~\bibnamefont
  {Wang}}, \bibinfo {author} {\bibfnamefont {W.}~\bibnamefont {Huang}},
  \bibinfo {author} {\bibfnamefont {F.}~\bibnamefont {Kondev}}, \bibinfo
  {author} {\bibfnamefont {G.}~\bibnamefont {Audi}},\ and\ \bibinfo {author}
  {\bibfnamefont {S.}~\bibnamefont {Naimi}},\ }\href
  {https://doi.org/10.1088/1674-1137/abddaf} {\bibfield  {journal} {\bibinfo
  {journal} {Chin. Phys. C}\ }\textbf {\bibinfo {volume} {45}},\ \bibinfo
  {pages} {030003} (\bibinfo {year} {2021})}\BibitemShut {NoStop}%
\bibitem [{\citenamefont {Kobayashi}\ \emph {et~al.}(2012)\citenamefont
  {Kobayashi}, \citenamefont {Nakamura}, \citenamefont {Tostevin},
  \citenamefont {Kondo}, \citenamefont {Aoi}, \citenamefont {Baba},
  \citenamefont {Deguchi}, \citenamefont {Gibelin}, \citenamefont {Ishihara},
  \citenamefont {Kawada}, \citenamefont {Kubo}, \citenamefont {Motobayashi},
  \citenamefont {Ohnishi}, \citenamefont {Orr}, \citenamefont {Otsu},
  \citenamefont {Sakurai}, \citenamefont {Satou}, \citenamefont {Simpson},
  \citenamefont {Sumikama}, \citenamefont {Takeda}, \citenamefont {Takechi},
  \citenamefont {Takeuchi}, \citenamefont {Tanaka}, \citenamefont {Tanaka},
  \citenamefont {Togano},\ and\ \citenamefont
  {Yoneda}}]{Kobayashi2012_PRC86-054604}%
  \BibitemOpen
  \bibfield  {author} {\bibinfo {author} {\bibfnamefont {N.}~\bibnamefont
  {Kobayashi}}, \bibinfo {author} {\bibfnamefont {T.}~\bibnamefont {Nakamura}},
  \bibinfo {author} {\bibfnamefont {J.~A.}\ \bibnamefont {Tostevin}}, \bibinfo
  {author} {\bibfnamefont {Y.}~\bibnamefont {Kondo}}, \bibinfo {author}
  {\bibfnamefont {N.}~\bibnamefont {Aoi}}, \bibinfo {author} {\bibfnamefont
  {H.}~\bibnamefont {Baba}}, \bibinfo {author} {\bibfnamefont {S.}~\bibnamefont
  {Deguchi}}, \bibinfo {author} {\bibfnamefont {J.}~\bibnamefont {Gibelin}},
  \bibinfo {author} {\bibfnamefont {M.}~\bibnamefont {Ishihara}}, \bibinfo
  {author} {\bibfnamefont {Y.}~\bibnamefont {Kawada}}, \bibinfo {author}
  {\bibfnamefont {T.}~\bibnamefont {Kubo}}, \bibinfo {author} {\bibfnamefont
  {T.}~\bibnamefont {Motobayashi}}, \bibinfo {author} {\bibfnamefont
  {T.}~\bibnamefont {Ohnishi}}, \bibinfo {author} {\bibfnamefont {N.~A.}\
  \bibnamefont {Orr}}, \bibinfo {author} {\bibfnamefont {H.}~\bibnamefont
  {Otsu}}, \bibinfo {author} {\bibfnamefont {H.}~\bibnamefont {Sakurai}},
  \bibinfo {author} {\bibfnamefont {Y.}~\bibnamefont {Satou}}, \bibinfo
  {author} {\bibfnamefont {E.~C.}\ \bibnamefont {Simpson}}, \bibinfo {author}
  {\bibfnamefont {T.}~\bibnamefont {Sumikama}}, \bibinfo {author}
  {\bibfnamefont {H.}~\bibnamefont {Takeda}}, \bibinfo {author} {\bibfnamefont
  {M.}~\bibnamefont {Takechi}}, \bibinfo {author} {\bibfnamefont
  {S.}~\bibnamefont {Takeuchi}}, \bibinfo {author} {\bibfnamefont {K.~N.}\
  \bibnamefont {Tanaka}}, \bibinfo {author} {\bibfnamefont {N.}~\bibnamefont
  {Tanaka}}, \bibinfo {author} {\bibfnamefont {Y.}~\bibnamefont {Togano}},\
  and\ \bibinfo {author} {\bibfnamefont {K.}~\bibnamefont {Yoneda}},\ }\href
  {https://doi.org/10.1103/PhysRevC.86.054604} {\bibfield  {journal} {\bibinfo
  {journal} {Phys. Rev. C}\ }\textbf {\bibinfo {volume} {86}},\ \bibinfo
  {pages} {054604} (\bibinfo {year} {2012})}\BibitemShut {NoStop}%
\bibitem [{\citenamefont {Horiuchi}\ and\ \citenamefont
  {Suzuki}(2006)}]{Horiuchi2006_PRC74-034311}%
  \BibitemOpen
  \bibfield  {author} {\bibinfo {author} {\bibfnamefont {W.}~\bibnamefont
  {Horiuchi}}\ and\ \bibinfo {author} {\bibfnamefont {Y.}~\bibnamefont
  {Suzuki}},\ }\href {https://doi.org/10.1103/PhysRevC.74.034311} {\bibfield
  {journal} {\bibinfo  {journal} {Phys. Rev. C}\ }\textbf {\bibinfo {volume}
  {74}},\ \bibinfo {pages} {034311} (\bibinfo {year} {2006})}\BibitemShut
  {NoStop}%
\bibitem [{\citenamefont {Ershov}\ \emph {et~al.}(2012)\citenamefont {Ershov},
  \citenamefont {Vaagen},\ and\ \citenamefont
  {Zhukov}}]{Ershov2012_PRC86-034331}%
  \BibitemOpen
  \bibfield  {author} {\bibinfo {author} {\bibfnamefont {S.~N.}\ \bibnamefont
  {Ershov}}, \bibinfo {author} {\bibfnamefont {J.~S.}\ \bibnamefont {Vaagen}},\
  and\ \bibinfo {author} {\bibfnamefont {M.~V.}\ \bibnamefont {Zhukov}},\
  }\href {https://doi.org/10.1103/PhysRevC.86.034331} {\bibfield  {journal}
  {\bibinfo  {journal} {Phys. Rev. C}\ }\textbf {\bibinfo {volume} {86}},\
  \bibinfo {pages} {034331} (\bibinfo {year} {2012})}\BibitemShut {NoStop}%
\bibitem [{\citenamefont {Pinilla}\ and\ \citenamefont
  {Descouvemont}(2016)}]{Pinilla2016_PRC94-024620}%
  \BibitemOpen
  \bibfield  {author} {\bibinfo {author} {\bibfnamefont {E.~C.}\ \bibnamefont
  {Pinilla}}\ and\ \bibinfo {author} {\bibfnamefont {P.}~\bibnamefont
  {Descouvemont}},\ }\href {https://doi.org/10.1103/PhysRevC.94.024620}
  {\bibfield  {journal} {\bibinfo  {journal} {Phys. Rev. C}\ }\textbf {\bibinfo
  {volume} {94}},\ \bibinfo {pages} {024620} (\bibinfo {year}
  {2016})}\BibitemShut {NoStop}%
\bibitem [{\citenamefont {Souza}\ \emph
  {et~al.}(2016{\natexlab{a}})\citenamefont {Souza}, \citenamefont {Garrido},\
  and\ \citenamefont {Frederico}}]{Souza2016_PRC94-064002}%
  \BibitemOpen
  \bibfield  {author} {\bibinfo {author} {\bibfnamefont {L.~A.}\ \bibnamefont
  {Souza}}, \bibinfo {author} {\bibfnamefont {E.}~\bibnamefont {Garrido}},\
  and\ \bibinfo {author} {\bibfnamefont {T.}~\bibnamefont {Frederico}},\ }\href
  {https://doi.org/10.1103/PhysRevC.94.064002} {\bibfield  {journal} {\bibinfo
  {journal} {Phys. Rev. C}\ }\textbf {\bibinfo {volume} {94}},\ \bibinfo
  {pages} {064002} (\bibinfo {year} {2016}{\natexlab{a}})}\BibitemShut
  {NoStop}%
\bibitem [{\citenamefont {Souza}\ \emph
  {et~al.}(2016{\natexlab{b}})\citenamefont {Souza}, \citenamefont {Bellotti},
  \citenamefont {Yamashita}, \citenamefont {Frederico},\ and\ \citenamefont
  {Tomio}}]{Souza2016_PLB757-368}%
  \BibitemOpen
  \bibfield  {author} {\bibinfo {author} {\bibfnamefont {L.}~\bibnamefont
  {Souza}}, \bibinfo {author} {\bibfnamefont {F.}~\bibnamefont {Bellotti}},
  \bibinfo {author} {\bibfnamefont {M.}~\bibnamefont {Yamashita}}, \bibinfo
  {author} {\bibfnamefont {T.}~\bibnamefont {Frederico}},\ and\ \bibinfo
  {author} {\bibfnamefont {L.}~\bibnamefont {Tomio}},\ }\href
  {https://doi.org/10.1016/j.physletb.2016.03.087} {\bibfield  {journal}
  {\bibinfo  {journal} {Phys. Lett. B}\ }\textbf {\bibinfo {volume} {757}},\
  \bibinfo {pages} {368} (\bibinfo {year} {2016}{\natexlab{b}})}\BibitemShut
  {NoStop}%
\bibitem [{\citenamefont {Shulgina}\ \emph {et~al.}(2018)\citenamefont
  {Shulgina}, \citenamefont {Ershov}, \citenamefont {Vaagen},\ and\
  \citenamefont {Zhukov}}]{Shulgina2018_PRC97-064307}%
  \BibitemOpen
  \bibfield  {author} {\bibinfo {author} {\bibfnamefont {N.~B.}\ \bibnamefont
  {Shulgina}}, \bibinfo {author} {\bibfnamefont {S.~N.}\ \bibnamefont
  {Ershov}}, \bibinfo {author} {\bibfnamefont {J.~S.}\ \bibnamefont {Vaagen}},\
  and\ \bibinfo {author} {\bibfnamefont {M.~V.}\ \bibnamefont {Zhukov}},\
  }\href {https://doi.org/10.1103/PhysRevC.97.064307} {\bibfield  {journal}
  {\bibinfo  {journal} {Phys. Rev. C}\ }\textbf {\bibinfo {volume} {97}},\
  \bibinfo {pages} {064307} (\bibinfo {year} {2018})}\BibitemShut {NoStop}%
\bibitem [{\citenamefont {Acharya}\ \emph {et~al.}(2013)\citenamefont
  {Acharya}, \citenamefont {Ji},\ and\ \citenamefont
  {Phillips}}]{Acharya2013_PLB723-196}%
  \BibitemOpen
  \bibfield  {author} {\bibinfo {author} {\bibfnamefont {B.}~\bibnamefont
  {Acharya}}, \bibinfo {author} {\bibfnamefont {C.}~\bibnamefont {Ji}},\ and\
  \bibinfo {author} {\bibfnamefont {D.}~\bibnamefont {Phillips}},\ }\href
  {https://doi.org/10.1016/j.physletb.2013.04.055} {\bibfield  {journal}
  {\bibinfo  {journal} {Phys. Lett. B}\ }\textbf {\bibinfo {volume} {723}},\
  \bibinfo {pages} {196} (\bibinfo {year} {2013})}\BibitemShut {NoStop}%
\bibitem [{\citenamefont {Hammer}\ \emph {et~al.}(2017)\citenamefont {Hammer},
  \citenamefont {Ji},\ and\ \citenamefont
  {Phillips}}]{Hammer2017_JPG44-103002}%
  \BibitemOpen
  \bibfield  {author} {\bibinfo {author} {\bibfnamefont {H.-W.}\ \bibnamefont
  {Hammer}}, \bibinfo {author} {\bibfnamefont {C.}~\bibnamefont {Ji}},\ and\
  \bibinfo {author} {\bibfnamefont {D.~R.}\ \bibnamefont {Phillips}},\ }\href
  {https://doi.org/10.1088/1361-6471/aa83db} {\bibfield  {journal} {\bibinfo
  {journal} {J. Phys. G: Nucl. Part. Phys.}\ }\textbf {\bibinfo {volume}
  {44}},\ \bibinfo {pages} {103002} (\bibinfo {year} {2017})}\BibitemShut
  {NoStop}%
\bibitem [{\citenamefont {Lu}\ \emph {et~al.}(2013)\citenamefont {Lu},
  \citenamefont {Sun},\ and\ \citenamefont {Long}}]{Lu2013_PRC87-034311}%
  \BibitemOpen
  \bibfield  {author} {\bibinfo {author} {\bibfnamefont {X.~L.}\ \bibnamefont
  {Lu}}, \bibinfo {author} {\bibfnamefont {B.~Y.}\ \bibnamefont {Sun}},\ and\
  \bibinfo {author} {\bibfnamefont {W.~H.}\ \bibnamefont {Long}},\ }\href
  {https://doi.org/10.1103/PhysRevC.87.034311} {\bibfield  {journal} {\bibinfo
  {journal} {Phys. Rev. C}\ }\textbf {\bibinfo {volume} {87}},\ \bibinfo
  {pages} {034311} (\bibinfo {year} {2013})}\BibitemShut {NoStop}%
\bibitem [{\citenamefont {Inakura}\ \emph {et~al.}(2014)\citenamefont
  {Inakura}, \citenamefont {Horiuchi}, \citenamefont {Suzuki},\ and\
  \citenamefont {Nakatsukasa}}]{Inakura2014_PRC89-064316}%
  \BibitemOpen
  \bibfield  {author} {\bibinfo {author} {\bibfnamefont {T.}~\bibnamefont
  {Inakura}}, \bibinfo {author} {\bibfnamefont {W.}~\bibnamefont {Horiuchi}},
  \bibinfo {author} {\bibfnamefont {Y.}~\bibnamefont {Suzuki}},\ and\ \bibinfo
  {author} {\bibfnamefont {T.}~\bibnamefont {Nakatsukasa}},\ }\href
  {https://doi.org/10.1103/PhysRevC.89.064316} {\bibfield  {journal} {\bibinfo
  {journal} {Phys. Rev. C}\ }\textbf {\bibinfo {volume} {89}},\ \bibinfo
  {pages} {064316} (\bibinfo {year} {2014})}\BibitemShut {NoStop}%
\bibitem [{\citenamefont {Hu}\ \emph {et~al.}(2019)\citenamefont {Hu},
  \citenamefont {Wu}, \citenamefont {Sun},\ and\ \citenamefont
  {Xu}}]{Hu2019_PRC99-061302R}%
  \BibitemOpen
  \bibfield  {author} {\bibinfo {author} {\bibfnamefont {B.~S.}\ \bibnamefont
  {Hu}}, \bibinfo {author} {\bibfnamefont {Q.}~\bibnamefont {Wu}}, \bibinfo
  {author} {\bibfnamefont {Z.~H.}\ \bibnamefont {Sun}},\ and\ \bibinfo {author}
  {\bibfnamefont {F.~R.}\ \bibnamefont {Xu}},\ }\href
  {https://doi.org/10.1103/PhysRevC.99.061302} {\bibfield  {journal} {\bibinfo
  {journal} {Phys. Rev. C}\ }\textbf {\bibinfo {volume} {99}},\ \bibinfo
  {pages} {061302(R)} (\bibinfo {year} {2019})}\BibitemShut {NoStop}%
\bibitem [{\citenamefont {Coraggio}\ \emph {et~al.}(2010)\citenamefont
  {Coraggio}, \citenamefont {Covello}, \citenamefont {Gargano},\ and\
  \citenamefont {Itaco}}]{Coraggio2010_PRC81-064303}%
  \BibitemOpen
  \bibfield  {author} {\bibinfo {author} {\bibfnamefont {L.}~\bibnamefont
  {Coraggio}}, \bibinfo {author} {\bibfnamefont {A.}~\bibnamefont {Covello}},
  \bibinfo {author} {\bibfnamefont {A.}~\bibnamefont {Gargano}},\ and\ \bibinfo
  {author} {\bibfnamefont {N.}~\bibnamefont {Itaco}},\ }\href
  {https://doi.org/10.1103/PhysRevC.81.064303} {\bibfield  {journal} {\bibinfo
  {journal} {Phys. Rev. C}\ }\textbf {\bibinfo {volume} {81}},\ \bibinfo
  {pages} {064303} (\bibinfo {year} {2010})}\BibitemShut {NoStop}%
\bibitem [{\citenamefont {Yuan}\ \emph {et~al.}(2012)\citenamefont {Yuan},
  \citenamefont {Suzuki}, \citenamefont {Otsuka}, \citenamefont {Xu},\ and\
  \citenamefont {Tsunoda}}]{Yuan2012_PRC85-064324}%
  \BibitemOpen
  \bibfield  {author} {\bibinfo {author} {\bibfnamefont {C.}~\bibnamefont
  {Yuan}}, \bibinfo {author} {\bibfnamefont {T.}~\bibnamefont {Suzuki}},
  \bibinfo {author} {\bibfnamefont {T.}~\bibnamefont {Otsuka}}, \bibinfo
  {author} {\bibfnamefont {F.}~\bibnamefont {Xu}},\ and\ \bibinfo {author}
  {\bibfnamefont {N.}~\bibnamefont {Tsunoda}},\ }\href
  {https://doi.org/10.1103/PhysRevC.85.064324} {\bibfield  {journal} {\bibinfo
  {journal} {Phys. Rev. C}\ }\textbf {\bibinfo {volume} {85}},\ \bibinfo
  {pages} {064324} (\bibinfo {year} {2012})}\BibitemShut {NoStop}%
\bibitem [{\citenamefont {Jansen}\ \emph {et~al.}(2014)\citenamefont {Jansen},
  \citenamefont {Engel}, \citenamefont {Hagen}, \citenamefont {Navratil},\ and\
  \citenamefont {Signoracci}}]{Jansen:2014qxa}%
  \BibitemOpen
  \bibfield  {author} {\bibinfo {author} {\bibfnamefont {G.}~\bibnamefont
  {Jansen}}, \bibinfo {author} {\bibfnamefont {J.}~\bibnamefont {Engel}},
  \bibinfo {author} {\bibfnamefont {G.}~\bibnamefont {Hagen}}, \bibinfo
  {author} {\bibfnamefont {P.}~\bibnamefont {Navratil}},\ and\ \bibinfo
  {author} {\bibfnamefont {A.}~\bibnamefont {Signoracci}},\ }\href
  {https://doi.org/10.1103/PhysRevLett.113.142502} {\bibfield  {journal}
  {\bibinfo  {journal} {Phys. Rev. Lett.}\ }\textbf {\bibinfo {volume} {113}},\
  \bibinfo {pages} {142502} (\bibinfo {year} {2014})},\ \Eprint
  {https://arxiv.org/abs/1402.2563} {arXiv:1402.2563 [nucl-th]} \BibitemShut
  {NoStop}%
\bibitem [{\citenamefont {Yao}\ \emph {et~al.}(2011)\citenamefont {Yao},
  \citenamefont {Meng}, \citenamefont {Ring}, \citenamefont {Li}, \citenamefont
  {Li},\ and\ \citenamefont {Hagino}}]{Yao2011_PRC84-024306}%
  \BibitemOpen
  \bibfield  {author} {\bibinfo {author} {\bibfnamefont {J.~M.}\ \bibnamefont
  {Yao}}, \bibinfo {author} {\bibfnamefont {J.}~\bibnamefont {Meng}}, \bibinfo
  {author} {\bibfnamefont {P.}~\bibnamefont {Ring}}, \bibinfo {author}
  {\bibfnamefont {Z.~X.}\ \bibnamefont {Li}}, \bibinfo {author} {\bibfnamefont
  {Z.~P.}\ \bibnamefont {Li}},\ and\ \bibinfo {author} {\bibfnamefont
  {K.}~\bibnamefont {Hagino}},\ }\href
  {https://doi.org/10.1103/PhysRevC.84.024306} {\bibfield  {journal} {\bibinfo
  {journal} {Phys. Rev. C}\ }\textbf {\bibinfo {volume} {84}},\ \bibinfo
  {pages} {024306} (\bibinfo {year} {2011})}\BibitemShut {NoStop}%
\bibitem [{\citenamefont {Li}\ \emph {et~al.}(2012)\citenamefont {Li},
  \citenamefont {Meng}, \citenamefont {Ring}, \citenamefont {Zhao},\ and\
  \citenamefont {Zhou}}]{Li2012_PRC85-024312}%
  \BibitemOpen
  \bibfield  {author} {\bibinfo {author} {\bibfnamefont {L.-L.}\ \bibnamefont
  {Li}}, \bibinfo {author} {\bibfnamefont {J.}~\bibnamefont {Meng}}, \bibinfo
  {author} {\bibfnamefont {P.}~\bibnamefont {Ring}}, \bibinfo {author}
  {\bibfnamefont {E.-G.}\ \bibnamefont {Zhao}},\ and\ \bibinfo {author}
  {\bibfnamefont {S.-G.}\ \bibnamefont {Zhou}},\ }\href
  {https://doi.org/10.1103/PhysRevC.85.024312} {\bibfield  {journal} {\bibinfo
  {journal} {Phys. Rev. C}\ }\textbf {\bibinfo {volume} {85}},\ \bibinfo
  {pages} {024312} (\bibinfo {year} {2012})}\BibitemShut {NoStop}%
\bibitem [{\citenamefont {Sun}\ and\ \citenamefont
  {Zhou}(2024)}]{Sun2024_NPR41-75}%
  \BibitemOpen
  \bibfield  {author} {\bibinfo {author} {\bibfnamefont {X.-X.}\ \bibnamefont
  {Sun}}\ and\ \bibinfo {author} {\bibfnamefont {S.-G.}\ \bibnamefont {Zhou}},\
  }\href {https://doi.org/10.11804/NuclPhysRev.41.2023CNPC56} {\bibfield
  {journal} {\bibinfo  {journal} {Nucl. Phys. Rev.}\ }\textbf {\bibinfo
  {volume} {41}},\ \bibinfo {pages} {75} (\bibinfo {year} {2024})}\BibitemShut
  {NoStop}%
\bibitem [{\citenamefont {Zhang}\ \emph {et~al.}(2025)\citenamefont {Zhang},
  \citenamefont {Pan}, \citenamefont {Wu}, \citenamefont {Qu}, \citenamefont
  {Lu},\ and\ \citenamefont {Sun}}]{Zhang:2025cxi}%
  \BibitemOpen
  \bibfield  {author} {\bibinfo {author} {\bibfnamefont {K.}~\bibnamefont
  {Zhang}}, \bibinfo {author} {\bibfnamefont {C.}~\bibnamefont {Pan}}, \bibinfo
  {author} {\bibfnamefont {X.}~\bibnamefont {Wu}}, \bibinfo {author}
  {\bibfnamefont {X.}~\bibnamefont {Qu}}, \bibinfo {author} {\bibfnamefont
  {X.}~\bibnamefont {Lu}},\ and\ \bibinfo {author} {\bibfnamefont
  {G.}~\bibnamefont {Sun}},\ }\href
  {https://doi.org/10.1007/s43673-025-00153-x} {\bibfield  {journal} {\bibinfo
  {journal} {AAPPS Bull.}\ }\textbf {\bibinfo {volume} {35}},\ \bibinfo {pages}
  {13} (\bibinfo {year} {2025})}\BibitemShut {NoStop}%
\bibitem [{\citenamefont {Zhang}\ \emph {et~al.}(2022)\citenamefont {Zhang},
  \citenamefont {Cheoun}, \citenamefont {Choi} \emph
  {et~al.}}]{DRHBcMassTable:2022uhi}%
  \BibitemOpen
  \bibfield  {author} {\bibinfo {author} {\bibfnamefont {K.}~\bibnamefont
  {Zhang}}, \bibinfo {author} {\bibfnamefont {M.-K.}\ \bibnamefont {Cheoun}},
  \bibinfo {author} {\bibfnamefont {Y.-B.}\ \bibnamefont {Choi}}, \emph
  {et~al.} (\bibinfo {collaboration} {DRHBc Mass Table}),\ }\href
  {https://doi.org/10.1016/j.adt.2022.101488} {\bibfield  {journal} {\bibinfo
  {journal} {Atom. Data Nucl. Data Tabl.}\ }\textbf {\bibinfo {volume} {144}},\
  \bibinfo {pages} {101488} (\bibinfo {year} {2022})},\ \Eprint
  {https://arxiv.org/abs/2201.03216} {arXiv:2201.03216 [nucl-th]} \BibitemShut
  {NoStop}%
\bibitem [{\citenamefont {Guo}\ \emph {et~al.}(2024)\citenamefont {Guo},
  \citenamefont {Cao}, \citenamefont {Chen}, \citenamefont {Chen},
  \citenamefont {Cheoun}, \citenamefont {Choi}, \citenamefont {Lam},
  \citenamefont {Deng}, \citenamefont {Dong}, \citenamefont {Du}, \citenamefont
  {Du}, \citenamefont {Duan}, \citenamefont {Fan}, \citenamefont {Gao},
  \citenamefont {Geng}, \citenamefont {Ha}, \citenamefont {He}, \citenamefont
  {Hu}, \citenamefont {Huang}, \citenamefont {Huang}, \citenamefont {Huang},
  \citenamefont {Huang}, \citenamefont {Hyung}, \citenamefont {Chan},
  \citenamefont {Jiang}, \citenamefont {Kim}, \citenamefont {Kim},
  \citenamefont {Lee}, \citenamefont {Lee}, \citenamefont {Li}, \citenamefont
  {Li}, \citenamefont {Li}, \citenamefont {Li}, \citenamefont {Lian},
  \citenamefont {Liang}, \citenamefont {Liu}, \citenamefont {Lu}, \citenamefont
  {Liu}, \citenamefont {Meng}, \citenamefont {Meng}, \citenamefont {Mun},
  \citenamefont {Niu}, \citenamefont {Niu}, \citenamefont {Pan}, \citenamefont
  {Peng}, \citenamefont {Qu}, \citenamefont {Papakonstantinou}, \citenamefont
  {Shang}, \citenamefont {Shang}, \citenamefont {Shen}, \citenamefont {Shen},
  \citenamefont {Sun}, \citenamefont {Sun}, \citenamefont {Wang}, \citenamefont
  {Wang}, \citenamefont {Wang}, \citenamefont {Wang}, \citenamefont {Wu},
  \citenamefont {Wu}, \citenamefont {Wu}, \citenamefont {Xia}, \citenamefont
  {Xie}, \citenamefont {Yao}, \citenamefont {Ip}, \citenamefont {Yiu},
  \citenamefont {Yu}, \citenamefont {Yu}, \citenamefont {Zhang}, \citenamefont
  {Zhang}, \citenamefont {Zhang}, \citenamefont {Zhang}, \citenamefont {Zhang},
  \citenamefont {Zhang}, \citenamefont {Zhang}, \citenamefont {Zhang},
  \citenamefont {Zhang}, \citenamefont {Zhao}, \citenamefont {Zhao},
  \citenamefont {Zheng}, \citenamefont {Zhou}, \citenamefont {Zhou},\ and\
  \citenamefont {Zou}}]{Guo2024_ADNDT158-101661}%
  \BibitemOpen
  \bibfield  {author} {\bibinfo {author} {\bibfnamefont {P.}~\bibnamefont
  {Guo}}, \bibinfo {author} {\bibfnamefont {X.}~\bibnamefont {Cao}}, \bibinfo
  {author} {\bibfnamefont {K.}~\bibnamefont {Chen}}, \bibinfo {author}
  {\bibfnamefont {Z.}~\bibnamefont {Chen}}, \bibinfo {author} {\bibfnamefont
  {M.-K.}\ \bibnamefont {Cheoun}}, \bibinfo {author} {\bibfnamefont {Y.-B.}\
  \bibnamefont {Choi}}, \bibinfo {author} {\bibfnamefont {P.~C.}\ \bibnamefont
  {Lam}}, \bibinfo {author} {\bibfnamefont {W.}~\bibnamefont {Deng}}, \bibinfo
  {author} {\bibfnamefont {J.}~\bibnamefont {Dong}}, \bibinfo {author}
  {\bibfnamefont {P.}~\bibnamefont {Du}}, \bibinfo {author} {\bibfnamefont
  {X.}~\bibnamefont {Du}}, \bibinfo {author} {\bibfnamefont {K.}~\bibnamefont
  {Duan}}, \bibinfo {author} {\bibfnamefont {X.}~\bibnamefont {Fan}}, \bibinfo
  {author} {\bibfnamefont {W.}~\bibnamefont {Gao}}, \bibinfo {author}
  {\bibfnamefont {L.}~\bibnamefont {Geng}}, \bibinfo {author} {\bibfnamefont
  {E.}~\bibnamefont {Ha}}, \bibinfo {author} {\bibfnamefont {X.-T.}\
  \bibnamefont {He}}, \bibinfo {author} {\bibfnamefont {J.}~\bibnamefont {Hu}},
  \bibinfo {author} {\bibfnamefont {J.}~\bibnamefont {Huang}}, \bibinfo
  {author} {\bibfnamefont {K.}~\bibnamefont {Huang}}, \bibinfo {author}
  {\bibfnamefont {Y.}~\bibnamefont {Huang}}, \bibinfo {author} {\bibfnamefont
  {Z.}~\bibnamefont {Huang}}, \bibinfo {author} {\bibfnamefont {K.~D.}\
  \bibnamefont {Hyung}}, \bibinfo {author} {\bibfnamefont {H.~Y.}\ \bibnamefont
  {Chan}}, \bibinfo {author} {\bibfnamefont {X.}~\bibnamefont {Jiang}},
  \bibinfo {author} {\bibfnamefont {S.}~\bibnamefont {Kim}}, \bibinfo {author}
  {\bibfnamefont {Y.}~\bibnamefont {Kim}}, \bibinfo {author} {\bibfnamefont
  {C.-H.}\ \bibnamefont {Lee}}, \bibinfo {author} {\bibfnamefont
  {J.}~\bibnamefont {Lee}}, \bibinfo {author} {\bibfnamefont {J.}~\bibnamefont
  {Li}}, \bibinfo {author} {\bibfnamefont {M.}~\bibnamefont {Li}}, \bibinfo
  {author} {\bibfnamefont {Z.}~\bibnamefont {Li}}, \bibinfo {author}
  {\bibfnamefont {Z.}~\bibnamefont {Li}}, \bibinfo {author} {\bibfnamefont
  {Z.}~\bibnamefont {Lian}}, \bibinfo {author} {\bibfnamefont {H.}~\bibnamefont
  {Liang}}, \bibinfo {author} {\bibfnamefont {L.}~\bibnamefont {Liu}}, \bibinfo
  {author} {\bibfnamefont {X.}~\bibnamefont {Lu}}, \bibinfo {author}
  {\bibfnamefont {Z.-R.}\ \bibnamefont {Liu}}, \bibinfo {author} {\bibfnamefont
  {J.}~\bibnamefont {Meng}}, \bibinfo {author} {\bibfnamefont {Z.}~\bibnamefont
  {Meng}}, \bibinfo {author} {\bibfnamefont {M.-H.}\ \bibnamefont {Mun}},
  \bibinfo {author} {\bibfnamefont {Y.}~\bibnamefont {Niu}}, \bibinfo {author}
  {\bibfnamefont {Z.}~\bibnamefont {Niu}}, \bibinfo {author} {\bibfnamefont
  {C.}~\bibnamefont {Pan}}, \bibinfo {author} {\bibfnamefont {J.}~\bibnamefont
  {Peng}}, \bibinfo {author} {\bibfnamefont {X.}~\bibnamefont {Qu}}, \bibinfo
  {author} {\bibfnamefont {P.}~\bibnamefont {Papakonstantinou}}, \bibinfo
  {author} {\bibfnamefont {T.}~\bibnamefont {Shang}}, \bibinfo {author}
  {\bibfnamefont {X.}~\bibnamefont {Shang}}, \bibinfo {author} {\bibfnamefont
  {C.}~\bibnamefont {Shen}}, \bibinfo {author} {\bibfnamefont {G.}~\bibnamefont
  {Shen}}, \bibinfo {author} {\bibfnamefont {T.}~\bibnamefont {Sun}}, \bibinfo
  {author} {\bibfnamefont {X.-X.}\ \bibnamefont {Sun}}, \bibinfo {author}
  {\bibfnamefont {S.}~\bibnamefont {Wang}}, \bibinfo {author} {\bibfnamefont
  {T.}~\bibnamefont {Wang}}, \bibinfo {author} {\bibfnamefont {Y.}~\bibnamefont
  {Wang}}, \bibinfo {author} {\bibfnamefont {Y.}~\bibnamefont {Wang}}, \bibinfo
  {author} {\bibfnamefont {J.}~\bibnamefont {Wu}}, \bibinfo {author}
  {\bibfnamefont {L.}~\bibnamefont {Wu}}, \bibinfo {author} {\bibfnamefont
  {X.}~\bibnamefont {Wu}}, \bibinfo {author} {\bibfnamefont {X.}~\bibnamefont
  {Xia}}, \bibinfo {author} {\bibfnamefont {H.}~\bibnamefont {Xie}}, \bibinfo
  {author} {\bibfnamefont {J.}~\bibnamefont {Yao}}, \bibinfo {author}
  {\bibfnamefont {K.~Y.}\ \bibnamefont {Ip}}, \bibinfo {author} {\bibfnamefont
  {T.~C.}\ \bibnamefont {Yiu}}, \bibinfo {author} {\bibfnamefont
  {J.}~\bibnamefont {Yu}}, \bibinfo {author} {\bibfnamefont {Y.}~\bibnamefont
  {Yu}}, \bibinfo {author} {\bibfnamefont {K.}~\bibnamefont {Zhang}}, \bibinfo
  {author} {\bibfnamefont {S.}~\bibnamefont {Zhang}}, \bibinfo {author}
  {\bibfnamefont {S.}~\bibnamefont {Zhang}}, \bibinfo {author} {\bibfnamefont
  {W.}~\bibnamefont {Zhang}}, \bibinfo {author} {\bibfnamefont
  {X.}~\bibnamefont {Zhang}}, \bibinfo {author} {\bibfnamefont
  {Y.}~\bibnamefont {Zhang}}, \bibinfo {author} {\bibfnamefont
  {Y.}~\bibnamefont {Zhang}}, \bibinfo {author} {\bibfnamefont
  {Y.}~\bibnamefont {Zhang}}, \bibinfo {author} {\bibfnamefont
  {Z.}~\bibnamefont {Zhang}}, \bibinfo {author} {\bibfnamefont
  {Q.}~\bibnamefont {Zhao}}, \bibinfo {author} {\bibfnamefont {Y.}~\bibnamefont
  {Zhao}}, \bibinfo {author} {\bibfnamefont {R.}~\bibnamefont {Zheng}},
  \bibinfo {author} {\bibfnamefont {C.}~\bibnamefont {Zhou}}, \bibinfo {author}
  {\bibfnamefont {S.-G.}\ \bibnamefont {Zhou}},\ and\ \bibinfo {author}
  {\bibfnamefont {L.}~\bibnamefont {Zou}},\ }\href
  {https://doi.org/10.1016/j.adt.2024.101661} {\bibfield  {journal} {\bibinfo
  {journal} {Atomic Data and Nuclear Data Tables}\ }\textbf {\bibinfo {volume}
  {158}},\ \bibinfo {pages} {101661} (\bibinfo {year} {2024})}\BibitemShut
  {NoStop}%
\bibitem [{\citenamefont {Ring}\ and\ \citenamefont {Schuck}(1980)}]{Ring1980}%
  \BibitemOpen
  \bibfield  {author} {\bibinfo {author} {\bibfnamefont {P.}~\bibnamefont
  {Ring}}\ and\ \bibinfo {author} {\bibfnamefont {P.}~\bibnamefont {Schuck}},\
  }\href@noop {} {\emph {\bibinfo {title} {The Nuclear Many-Body Problem}}}\
  (\bibinfo  {publisher} {Springer-Verlag Berlin Heidelberg},\ \bibinfo {year}
  {1980})\BibitemShut {NoStop}%
\bibitem [{\citenamefont {Egido}(2016)}]{Egido2016_PS91-073003}%
  \BibitemOpen
  \bibfield  {author} {\bibinfo {author} {\bibfnamefont {J.~L.}\ \bibnamefont
  {Egido}},\ }\href {https://doi.org/10.1088/0031-8949/91/7/073003} {\bibfield
  {journal} {\bibinfo  {journal} {Phys. Scr.}\ }\textbf {\bibinfo {volume}
  {91}},\ \bibinfo {pages} {073003} (\bibinfo {year} {2016})}\BibitemShut
  {NoStop}%
\bibitem [{\citenamefont {Nik{\v s}i{\'c}}\ \emph {et~al.}(2006)\citenamefont
  {Nik{\v s}i{\'c}}, \citenamefont {Vretenar},\ and\ \citenamefont
  {Ring}}]{Niksic2006_PRC73-034308}%
  \BibitemOpen
  \bibfield  {author} {\bibinfo {author} {\bibfnamefont {T.}~\bibnamefont
  {Nik{\v s}i{\'c}}}, \bibinfo {author} {\bibfnamefont {D.}~\bibnamefont
  {Vretenar}},\ and\ \bibinfo {author} {\bibfnamefont {P.}~\bibnamefont
  {Ring}},\ }\href {https://doi.org/10.1103/PhysRevC.73.034308} {\bibfield
  {journal} {\bibinfo  {journal} {Phys. Rev. C}\ }\textbf {\bibinfo {volume}
  {73}},\ \bibinfo {pages} {034308} (\bibinfo {year} {2006})}\BibitemShut
  {NoStop}%
\bibitem [{\citenamefont {Yao}\ \emph {et~al.}(2010)\citenamefont {Yao},
  \citenamefont {Meng}, \citenamefont {Ring},\ and\ \citenamefont
  {Vretenar}}]{Yao2010_PRC81-044311}%
  \BibitemOpen
  \bibfield  {author} {\bibinfo {author} {\bibfnamefont {J.~M.}\ \bibnamefont
  {Yao}}, \bibinfo {author} {\bibfnamefont {J.}~\bibnamefont {Meng}}, \bibinfo
  {author} {\bibfnamefont {P.}~\bibnamefont {Ring}},\ and\ \bibinfo {author}
  {\bibfnamefont {D.}~\bibnamefont {Vretenar}},\ }\href
  {https://doi.org/10.1103/PhysRevC.81.044311} {\bibfield  {journal} {\bibinfo
  {journal} {Phys. Rev. C}\ }\textbf {\bibinfo {volume} {81}},\ \bibinfo
  {pages} {044311} (\bibinfo {year} {2010})}\BibitemShut {NoStop}%
\bibitem [{\citenamefont {Nik{\v s}i{\'c}}\ \emph {et~al.}(2011)\citenamefont
  {Nik{\v s}i{\'c}}, \citenamefont {Vretenar},\ and\ \citenamefont
  {Ring}}]{Niksic2011_PPNP66-519}%
  \BibitemOpen
  \bibfield  {author} {\bibinfo {author} {\bibfnamefont {T.}~\bibnamefont
  {Nik{\v s}i{\'c}}}, \bibinfo {author} {\bibfnamefont {D.}~\bibnamefont
  {Vretenar}},\ and\ \bibinfo {author} {\bibfnamefont {P.}~\bibnamefont
  {Ring}},\ }\href {https://doi.org/10.1016/j.ppnp.2011.01.055} {\bibfield
  {journal} {\bibinfo  {journal} {Prog. Part. Nucl. Phys.}\ }\textbf {\bibinfo
  {volume} {66}},\ \bibinfo {pages} {519} (\bibinfo {year} {2011})}\BibitemShut
  {NoStop}%
\bibitem [{\citenamefont {Zhou}\ \emph {et~al.}(2003)\citenamefont {Zhou},
  \citenamefont {Meng},\ and\ \citenamefont {Ring}}]{Zhou2003_PRC68-034323}%
  \BibitemOpen
  \bibfield  {author} {\bibinfo {author} {\bibfnamefont {S.-G.}\ \bibnamefont
  {Zhou}}, \bibinfo {author} {\bibfnamefont {J.}~\bibnamefont {Meng}},\ and\
  \bibinfo {author} {\bibfnamefont {P.}~\bibnamefont {Ring}},\ }\href
  {https://doi.org/10.1103/PhysRevC.68.034323} {\bibfield  {journal} {\bibinfo
  {journal} {Phys. Rev. C}\ }\textbf {\bibinfo {volume} {68}},\ \bibinfo
  {pages} {034323} (\bibinfo {year} {2003})}\BibitemShut {NoStop}%
\bibitem [{\citenamefont {Sun}\ and\ \citenamefont
  {Zhou}(2021{\natexlab{a}})}]{Sun2021_SciBull66-1521}%
  \BibitemOpen
  \bibfield  {author} {\bibinfo {author} {\bibfnamefont {X.-X.}\ \bibnamefont
  {Sun}}\ and\ \bibinfo {author} {\bibfnamefont {S.-G.}\ \bibnamefont {Zhou}},\
  }\href {https://doi.org/10.1016/j.scib.2021.07.005} {\bibfield  {journal}
  {\bibinfo  {journal} {Sci. Bull.}\ }\textbf {\bibinfo {volume} {66}},\
  \bibinfo {pages} {1521} (\bibinfo {year} {2021}{\natexlab{a}})},\ \Eprint
  {https://arxiv.org/abs/2103.10886} {arXiv:2103.10886 [nucl-th]} \BibitemShut
  {NoStop}%
\bibitem [{\citenamefont {Sun}\ and\ \citenamefont
  {Zhou}(2021{\natexlab{b}})}]{Sun2021_PRC104-064319}%
  \BibitemOpen
  \bibfield  {author} {\bibinfo {author} {\bibfnamefont {X.-X.}\ \bibnamefont
  {Sun}}\ and\ \bibinfo {author} {\bibfnamefont {S.-G.}\ \bibnamefont {Zhou}},\
  }\href {https://doi.org/10.1103/PhysRevC.104.064319} {\bibfield  {journal}
  {\bibinfo  {journal} {Phys. Rev. C}\ }\textbf {\bibinfo {volume} {104}},\
  \bibinfo {pages} {064319} (\bibinfo {year} {2021}{\natexlab{b}})}\BibitemShut
  {NoStop}%
\bibitem [{\citenamefont {Zhao}\ \emph {et~al.}(2010)\citenamefont {Zhao},
  \citenamefont {Li}, \citenamefont {Yao},\ and\ \citenamefont
  {Meng}}]{Zhao2010_PRC82-054319}%
  \BibitemOpen
  \bibfield  {author} {\bibinfo {author} {\bibfnamefont {P.~W.}\ \bibnamefont
  {Zhao}}, \bibinfo {author} {\bibfnamefont {Z.~P.}\ \bibnamefont {Li}},
  \bibinfo {author} {\bibfnamefont {J.~M.}\ \bibnamefont {Yao}},\ and\ \bibinfo
  {author} {\bibfnamefont {J.}~\bibnamefont {Meng}},\ }\href
  {https://doi.org/10.1103/PhysRevC.82.054319} {\bibfield  {journal} {\bibinfo
  {journal} {Phys. Rev. C}\ }\textbf {\bibinfo {volume} {82}},\ \bibinfo
  {pages} {054319} (\bibinfo {year} {2010})}\BibitemShut {NoStop}%
\bibitem [{\citenamefont {B{\"u}rvenich}\ \emph {et~al.}(2002)\citenamefont
  {B{\"u}rvenich}, \citenamefont {Madland}, \citenamefont {Maruhn},\ and\
  \citenamefont {Reinhard}}]{Burvenich2002_PRC65-044308}%
  \BibitemOpen
  \bibfield  {author} {\bibinfo {author} {\bibfnamefont {T.}~\bibnamefont
  {B{\"u}rvenich}}, \bibinfo {author} {\bibfnamefont {D.~G.}\ \bibnamefont
  {Madland}}, \bibinfo {author} {\bibfnamefont {J.~A.}\ \bibnamefont
  {Maruhn}},\ and\ \bibinfo {author} {\bibfnamefont {P.-G.}\ \bibnamefont
  {Reinhard}},\ }\href {https://doi.org/10.1103/PhysRevC.65.044308} {\bibfield
  {journal} {\bibinfo  {journal} {Phys. Rev. C}\ }\textbf {\bibinfo {volume}
  {65}},\ \bibinfo {pages} {044308} (\bibinfo {year} {2002})}\BibitemShut
  {NoStop}%
\bibitem [{\citenamefont {Nik{\v s}i{\'c}}\ \emph {et~al.}(2008)\citenamefont
  {Nik{\v s}i{\'c}}, \citenamefont {Vretenar},\ and\ \citenamefont
  {Ring}}]{Niksic2008_PRC78-034318}%
  \BibitemOpen
  \bibfield  {author} {\bibinfo {author} {\bibfnamefont {T.}~\bibnamefont
  {Nik{\v s}i{\'c}}}, \bibinfo {author} {\bibfnamefont {D.}~\bibnamefont
  {Vretenar}},\ and\ \bibinfo {author} {\bibfnamefont {P.}~\bibnamefont
  {Ring}},\ }\href {https://doi.org/10.1103/PhysRevC.78.034318} {\bibfield
  {journal} {\bibinfo  {journal} {Phys. Rev. C}\ }\textbf {\bibinfo {volume}
  {78}},\ \bibinfo {pages} {034318} (\bibinfo {year} {2008})}\BibitemShut
  {NoStop}%
\bibitem [{\citenamefont {Zhang}\ \emph {et~al.}(2020)\citenamefont {Zhang},
  \citenamefont {Cheoun}, \citenamefont {Choi}, \citenamefont {Chong},
  \citenamefont {Dong}, \citenamefont {Geng}, \citenamefont {Ha}, \citenamefont
  {He}, \citenamefont {Heo}, \citenamefont {Ho}, \citenamefont {In},
  \citenamefont {Kim}, \citenamefont {Kim}, \citenamefont {Lee}, \citenamefont
  {Lee}, \citenamefont {Li}, \citenamefont {Luo}, \citenamefont {Meng},
  \citenamefont {Mun}, \citenamefont {Niu}, \citenamefont {Pan}, \citenamefont
  {Papakonstantinou}, \citenamefont {Shang}, \citenamefont {Shen},
  \citenamefont {Shen}, \citenamefont {Sun}, \citenamefont {Sun}, \citenamefont
  {Tam}, \citenamefont {{Thaivayongnou}}, \citenamefont {Wang}, \citenamefont
  {Wong}, \citenamefont {Xia}, \citenamefont {Yan}, \citenamefont {Yeung},
  \citenamefont {Yiu}, \citenamefont {Zhang}, \citenamefont {Zhang},\ and\
  \citenamefont {Zhou}}]{Zhang2020_PRC102-024314}%
  \BibitemOpen
  \bibfield  {author} {\bibinfo {author} {\bibfnamefont {K.}~\bibnamefont
  {Zhang}}, \bibinfo {author} {\bibfnamefont {M.-K.}\ \bibnamefont {Cheoun}},
  \bibinfo {author} {\bibfnamefont {Y.-B.}\ \bibnamefont {Choi}}, \bibinfo
  {author} {\bibfnamefont {P.~S.}\ \bibnamefont {Chong}}, \bibinfo {author}
  {\bibfnamefont {J.}~\bibnamefont {Dong}}, \bibinfo {author} {\bibfnamefont
  {L.}~\bibnamefont {Geng}}, \bibinfo {author} {\bibfnamefont {E.}~\bibnamefont
  {Ha}}, \bibinfo {author} {\bibfnamefont {X.}~\bibnamefont {He}}, \bibinfo
  {author} {\bibfnamefont {C.}~\bibnamefont {Heo}}, \bibinfo {author}
  {\bibfnamefont {M.~C.}\ \bibnamefont {Ho}}, \bibinfo {author} {\bibfnamefont
  {E.~J.}\ \bibnamefont {In}}, \bibinfo {author} {\bibfnamefont
  {S.}~\bibnamefont {Kim}}, \bibinfo {author} {\bibfnamefont {Y.}~\bibnamefont
  {Kim}}, \bibinfo {author} {\bibfnamefont {C.-H.}\ \bibnamefont {Lee}},
  \bibinfo {author} {\bibfnamefont {J.}~\bibnamefont {Lee}}, \bibinfo {author}
  {\bibfnamefont {Z.}~\bibnamefont {Li}}, \bibinfo {author} {\bibfnamefont
  {T.}~\bibnamefont {Luo}}, \bibinfo {author} {\bibfnamefont {J.}~\bibnamefont
  {Meng}}, \bibinfo {author} {\bibfnamefont {M.-H.}\ \bibnamefont {Mun}},
  \bibinfo {author} {\bibfnamefont {Z.}~\bibnamefont {Niu}}, \bibinfo {author}
  {\bibfnamefont {C.}~\bibnamefont {Pan}}, \bibinfo {author} {\bibfnamefont
  {P.}~\bibnamefont {Papakonstantinou}}, \bibinfo {author} {\bibfnamefont
  {X.}~\bibnamefont {Shang}}, \bibinfo {author} {\bibfnamefont
  {C.}~\bibnamefont {Shen}}, \bibinfo {author} {\bibfnamefont {G.}~\bibnamefont
  {Shen}}, \bibinfo {author} {\bibfnamefont {W.}~\bibnamefont {Sun}}, \bibinfo
  {author} {\bibfnamefont {X.-X.}\ \bibnamefont {Sun}}, \bibinfo {author}
  {\bibfnamefont {C.~K.}\ \bibnamefont {Tam}}, \bibinfo {author} {\bibnamefont
  {{Thaivayongnou}}}, \bibinfo {author} {\bibfnamefont {C.}~\bibnamefont
  {Wang}}, \bibinfo {author} {\bibfnamefont {S.~H.}\ \bibnamefont {Wong}},
  \bibinfo {author} {\bibfnamefont {X.}~\bibnamefont {Xia}}, \bibinfo {author}
  {\bibfnamefont {Y.}~\bibnamefont {Yan}}, \bibinfo {author} {\bibfnamefont
  {R.~W.-Y.}\ \bibnamefont {Yeung}}, \bibinfo {author} {\bibfnamefont {T.~C.}\
  \bibnamefont {Yiu}}, \bibinfo {author} {\bibfnamefont {S.}~\bibnamefont
  {Zhang}}, \bibinfo {author} {\bibfnamefont {W.}~\bibnamefont {Zhang}},\ and\
  \bibinfo {author} {\bibfnamefont {S.-G.}\ \bibnamefont {Zhou}},\ }\href
  {https://doi.org/10.1103/physrevc.102.024314} {\bibfield  {journal} {\bibinfo
   {journal} {Phys. Rev. C}\ }\textbf {\bibinfo {volume} {102}},\ \bibinfo
  {pages} {024314} (\bibinfo {year} {2020})}\BibitemShut {NoStop}%
\bibitem [{\citenamefont {Robledo}(2009)}]{Robledo2009_PRC79-021302}%
  \BibitemOpen
  \bibfield  {author} {\bibinfo {author} {\bibfnamefont {L.~M.}\ \bibnamefont
  {Robledo}},\ }\href {https://doi.org/10.1103/PhysRevC.79.021302} {\bibfield
  {journal} {\bibinfo  {journal} {Phys. Rev. C}\ }\textbf {\bibinfo {volume}
  {79}},\ \bibinfo {pages} {021302} (\bibinfo {year} {2009})}\BibitemShut
  {NoStop}%
\bibitem [{\citenamefont {Fomenko}(1970)}]{Fomenko1970_JPA3-8}%
  \BibitemOpen
  \bibfield  {author} {\bibinfo {author} {\bibfnamefont {V.~N.}\ \bibnamefont
  {Fomenko}},\ }\href@noop {} {\bibfield  {journal} {\bibinfo  {journal} {J.
  Phys. A: Gen. Phys.}\ }\textbf {\bibinfo {volume} {3}},\ \bibinfo {pages} {8}
  (\bibinfo {year} {1970})}\BibitemShut {NoStop}%
\bibitem [{\citenamefont {Yao}(2022)}]{Yao2022_HNP}%
  \BibitemOpen
  \bibfield  {author} {\bibinfo {author} {\bibfnamefont {J.~M.}\ \bibnamefont
  {Yao}},\ }in\ \href {https://doi.org/10.1007/978-981-15-8818-1_18-1} {\emph
  {\bibinfo {booktitle} {Handbook of Nuclear Physics}}},\ \bibinfo {editor}
  {edited by\ \bibinfo {editor} {\bibfnamefont {I.}~\bibnamefont {Tanihata}},
  \bibinfo {editor} {\bibfnamefont {H.}~\bibnamefont {Toki}},\ and\ \bibinfo
  {editor} {\bibfnamefont {T.}~\bibnamefont {Kajino}}}\ (\bibinfo  {publisher}
  {Springer Nature Singapore},\ \bibinfo {address} {Singapore},\ \bibinfo
  {year} {2022})\ pp.\ \bibinfo {pages} {1--36}\BibitemShut {NoStop}%
\bibitem [{\citenamefont {Zhou}\ and\ \citenamefont
  {Yao}(2023)}]{Zhou2023_IJMPE32-2340011}%
  \BibitemOpen
  \bibfield  {author} {\bibinfo {author} {\bibfnamefont {E.}~\bibnamefont
  {Zhou}}\ and\ \bibinfo {author} {\bibfnamefont {J.}~\bibnamefont {Yao}},\
  }\href {https://doi.org/10.1142/S0218301323400116} {\bibfield  {journal}
  {\bibinfo  {journal} {Int. J. Mod. Phys. E}\ }\textbf {\bibinfo {volume}
  {32}},\ \bibinfo {pages} {2340011} (\bibinfo {year} {2023})},\ \Eprint
  {https://arxiv.org/abs/2309.09488} {arXiv:2309.09488 [nucl-th]} \BibitemShut
  {NoStop}%
\bibitem [{\citenamefont {Meng}(1998)}]{Meng1998_NPA635-3}%
  \BibitemOpen
  \bibfield  {author} {\bibinfo {author} {\bibfnamefont {J.}~\bibnamefont
  {Meng}},\ }\href {https://doi.org/10.1016/S0375-9474(98)00178-X} {\bibfield
  {journal} {\bibinfo  {journal} {Nucl. Phys. A}\ }\textbf {\bibinfo {volume}
  {635}},\ \bibinfo {pages} {3} (\bibinfo {year} {1998})}\BibitemShut {NoStop}%
\bibitem [{\citenamefont {Sun}\ \emph {et~al.}(2020)\citenamefont {Sun},
  \citenamefont {Zhao},\ and\ \citenamefont {Zhou}}]{Sun2020_NPA1003-122011}%
  \BibitemOpen
  \bibfield  {author} {\bibinfo {author} {\bibfnamefont {X.-X.}\ \bibnamefont
  {Sun}}, \bibinfo {author} {\bibfnamefont {J.}~\bibnamefont {Zhao}},\ and\
  \bibinfo {author} {\bibfnamefont {S.-G.}\ \bibnamefont {Zhou}},\ }\href
  {https://doi.org/10.1016/j.nuclphysa.2020.122011} {\bibfield  {journal}
  {\bibinfo  {journal} {Nucl. Phys. A}\ }\textbf {\bibinfo {volume} {1003}},\
  \bibinfo {pages} {122011} (\bibinfo {year} {2020})},\ \Eprint
  {https://arxiv.org/abs/2008.04664} {arXiv:2008.04664 [nucl-th]} \BibitemShut
  {NoStop}%
\bibitem [{\citenamefont {Long}\ \emph {et~al.}(2004)\citenamefont {Long},
  \citenamefont {Meng}, \citenamefont {Giai},\ and\ \citenamefont
  {Zhou}}]{Long2004_PRC69-034319}%
  \BibitemOpen
  \bibfield  {author} {\bibinfo {author} {\bibfnamefont {W.}~\bibnamefont
  {Long}}, \bibinfo {author} {\bibfnamefont {J.}~\bibnamefont {Meng}}, \bibinfo
  {author} {\bibfnamefont {N.~V.}\ \bibnamefont {Giai}},\ and\ \bibinfo
  {author} {\bibfnamefont {S.-G.}\ \bibnamefont {Zhou}},\ }\href
  {https://doi.org/10.1103/PhysRevC.69.034319} {\bibfield  {journal} {\bibinfo
  {journal} {Phys. Rev. C}\ }\textbf {\bibinfo {volume} {69}},\ \bibinfo
  {pages} {034319} (\bibinfo {year} {2004})}\BibitemShut {NoStop}%
\bibitem [{\citenamefont {Toki}\ \emph {et~al.}(1995)\citenamefont {Toki},
  \citenamefont {Hirata}, \citenamefont {Sugahara}, \citenamefont {Sumiyoshi},\
  and\ \citenamefont {Tanihata}}]{Toki1995_NPA588-c357}%
  \BibitemOpen
  \bibfield  {author} {\bibinfo {author} {\bibfnamefont {H.}~\bibnamefont
  {Toki}}, \bibinfo {author} {\bibfnamefont {D.}~\bibnamefont {Hirata}},
  \bibinfo {author} {\bibfnamefont {Y.}~\bibnamefont {Sugahara}}, \bibinfo
  {author} {\bibfnamefont {K.}~\bibnamefont {Sumiyoshi}},\ and\ \bibinfo
  {author} {\bibfnamefont {I.}~\bibnamefont {Tanihata}},\ }\href
  {https://doi.org/10.1016/0375-9474(95)00161-S} {\bibfield  {journal}
  {\bibinfo  {journal} {Nucl. Phys. A}\ }\textbf {\bibinfo {volume} {588}},\
  \bibinfo {pages} {c357} (\bibinfo {year} {1995})}\BibitemShut {NoStop}%
\bibitem [{\citenamefont {Nik{\v s}i{\'c}}\ \emph {et~al.}(2007)\citenamefont
  {Nik{\v s}i{\'c}}, \citenamefont {Vretenar}, \citenamefont {Lalazissis},\
  and\ \citenamefont {Ring}}]{Niksic2007_PRL99-092502}%
  \BibitemOpen
  \bibfield  {author} {\bibinfo {author} {\bibfnamefont {T.}~\bibnamefont
  {Nik{\v s}i{\'c}}}, \bibinfo {author} {\bibfnamefont {D.}~\bibnamefont
  {Vretenar}}, \bibinfo {author} {\bibfnamefont {G.~A.}\ \bibnamefont
  {Lalazissis}},\ and\ \bibinfo {author} {\bibfnamefont {P.}~\bibnamefont
  {Ring}},\ }\href {https://doi.org/10.1103/PhysRevLett.99.092502} {\bibfield
  {journal} {\bibinfo  {journal} {Phys. Rev. Lett.}\ }\textbf {\bibinfo
  {volume} {99}},\ \bibinfo {pages} {092502} (\bibinfo {year}
  {2007})}\BibitemShut {NoStop}%
\bibitem [{\citenamefont {Hagino}\ and\ \citenamefont
  {Sagawa}(2005)}]{Hagino2005_PRC72-44321}%
  \BibitemOpen
  \bibfield  {author} {\bibinfo {author} {\bibfnamefont {K.}~\bibnamefont
  {Hagino}}\ and\ \bibinfo {author} {\bibfnamefont {H.}~\bibnamefont
  {Sagawa}},\ }\href {https://doi.org/10.1103/PhysRevC.72.044321} {\bibfield
  {journal} {\bibinfo  {journal} {Phys. Rev. C}\ }\textbf {\bibinfo {volume}
  {72}},\ \bibinfo {pages} {044321} (\bibinfo {year} {2005})}\BibitemShut
  {NoStop}%
\bibitem [{\citenamefont {Matsuo}\ \emph {et~al.}(2005)\citenamefont {Matsuo},
  \citenamefont {Mizuyama},\ and\ \citenamefont
  {Serizawa}}]{Matsuo2005_PRC71-064326}%
  \BibitemOpen
  \bibfield  {author} {\bibinfo {author} {\bibfnamefont {M.}~\bibnamefont
  {Matsuo}}, \bibinfo {author} {\bibfnamefont {K.}~\bibnamefont {Mizuyama}},\
  and\ \bibinfo {author} {\bibfnamefont {Y.}~\bibnamefont {Serizawa}},\ }\href
  {https://doi.org/10.1103/physrevc.71.064326} {\bibfield  {journal} {\bibinfo
  {journal} {Phys. Rev. C}\ }\textbf {\bibinfo {volume} {71}},\ \bibinfo
  {pages} {064326} (\bibinfo {year} {2005})}\BibitemShut {NoStop}%
\bibitem [{\citenamefont {Navr{\'a}til}\ \emph {et~al.}(2026)\citenamefont
  {Navr{\'a}til}, \citenamefont {Quaglioni}, \citenamefont {Hupin},
  \citenamefont {Gennari},\ and\ \citenamefont
  {Kravvaris}}]{Navratil2026_Particle9-57}%
  \BibitemOpen
  \bibfield  {author} {\bibinfo {author} {\bibfnamefont {P.}~\bibnamefont
  {Navr{\'a}til}}, \bibinfo {author} {\bibfnamefont {S.}~\bibnamefont
  {Quaglioni}}, \bibinfo {author} {\bibfnamefont {G.}~\bibnamefont {Hupin}},
  \bibinfo {author} {\bibfnamefont {M.}~\bibnamefont {Gennari}},\ and\ \bibinfo
  {author} {\bibfnamefont {K.}~\bibnamefont {Kravvaris}},\ }\href
  {https://doi.org/10.3390/particles9020057} {\bibfield  {journal} {\bibinfo
  {journal} {Particles}\ }\textbf {\bibinfo {volume} {9}},\ \bibinfo {pages}
  {57} (\bibinfo {year} {2026})}\BibitemShut {NoStop}%
\bibitem [{\citenamefont {Zhang}\ \emph {et~al.}(2026)\citenamefont {Zhang},
  \citenamefont {Elhatisari},\ and\ \citenamefont {Mei{\ss}ner}}]{Zhang2026}%
  \BibitemOpen
  \bibfield  {author} {\bibinfo {author} {\bibfnamefont {S.}~\bibnamefont
  {Zhang}}, \bibinfo {author} {\bibfnamefont {S.}~\bibnamefont {Elhatisari}},\
  and\ \bibinfo {author} {\bibfnamefont {U.-G.}\ \bibnamefont {Mei{\ss}ner}},\
  }\href {https://doi.org/10.48550/arXiv.2512.18849} {\bibinfo {title}
  {Multi-neutron correlations in light nuclei via ab-initio lattice
  simulations}} (\bibinfo {year} {2026}),\ \Eprint
  {https://arxiv.org/abs/2512.18849} {arXiv:2512.18849 [nucl-th]} \BibitemShut
  {NoStop}%
\bibitem [{\citenamefont {Yang}\ and\ \citenamefont {Zhao}(2026)}]{Yang2026}%
  \BibitemOpen
  \bibfield  {author} {\bibinfo {author} {\bibfnamefont {Y.}~\bibnamefont
  {Yang}}\ and\ \bibinfo {author} {\bibfnamefont {P.}~\bibnamefont {Zhao}},\
  }\href {https://doi.org/10.48550/arXiv.2607.24636} {\bibinfo {title}
  {Emergence of the halo in \$\textasciicircum\textbraceleft
  11\textbraceright\${{Li}} from full nuclear many-body dynamics}} (\bibinfo
  {year} {2026}),\ \Eprint {https://arxiv.org/abs/2607.24636} {arXiv:2607.24636
  [nucl-th]} \BibitemShut {NoStop}%
\bibitem [{\citenamefont {Alhassid}\ \emph {et~al.}(1982)\citenamefont
  {Alhassid}, \citenamefont {Gai},\ and\ \citenamefont
  {Bertsch}}]{Alhassid1982_PRL49-1482}%
  \BibitemOpen
  \bibfield  {author} {\bibinfo {author} {\bibfnamefont {Y.}~\bibnamefont
  {Alhassid}}, \bibinfo {author} {\bibfnamefont {M.}~\bibnamefont {Gai}},\ and\
  \bibinfo {author} {\bibfnamefont {G.~F.}\ \bibnamefont {Bertsch}},\ }\href
  {https://doi.org/10.1103/PhysRevLett.49.1482} {\bibfield  {journal} {\bibinfo
   {journal} {Phys. Rev. Lett.}\ }\textbf {\bibinfo {volume} {49}},\ \bibinfo
  {pages} {1482} (\bibinfo {year} {1982})}\BibitemShut {NoStop}%
\end{thebibliography}
%

\clearpage
\onecolumngrid
\setcounter{section}{0}
\setcounter{equation}{0}
\setcounter{figure}{0}
\setcounter{table}{0}
\renewcommand{\thesection}{S\arabic{section}}
\renewcommand{\theequation}{S\arabic{equation}}
\renewcommand{\thefigure}{S\arabic{figure}}
\renewcommand{\thetable}{S\arabic{table}}

\makeatletter
\renewcommand{\c@secnumdepth}{0}
\makeatother
\section*{Supplemental Material:
Emergence of the nuclear halo in 
$^{22}$C: bridging shape and shell}
\label{sec:supp}

\subsection{Numerical details of the MR-DRHBc calculations}
\label{sec:supp1}

\subsubsection{DRHBc mean field calculations}

The intrinsic states are axially deformed relativistic Hartree-Bogoliubov
vacua generated with the DRHBc theory~\cite{Zhou2010_PRC82-011301R,Li2012_PRC85-024312} in a
Dirac Woods--Saxon basis~\cite{Zhou2003_PRC68-034323}.
The radial box is $r_{\max}=20$~fm with mesh step $0.1$~fm. 
The Woods--Saxon cutoff is
$E_{\mathrm{cut}}=200$~MeV. 
Three point-coupling energy density functionals are used: PC-PK1 \cite{Zhao2010_PRC82-054319},
PC-F1 \cite{Burvenich2002_PRC65-044308}, and
DD-PC1 \cite{Niksic2008_PRC78-034318}. 
A zero-range
density-dependent pairing force~\cite{Meng1998_NPA635-3} with a sharp energy cutoff
of $100$~MeV is employed in the particle-particle channel,
\begin{equation}
  V^{\mathrm{pp}}(\mathbf{r}_1,\mathbf{r}_2)
  = \frac{V_{0}}{2}(1-P^{\sigma})
    \delta(\mathbf{r}_1-\mathbf{r}_2)
    \left(1-\frac{\rho(\mathbf{r}_1)}{\rho_{\mathrm{sat}}}\right),
\end{equation}
where $\rho_{\mathrm{sat}}$ is the saturation density of the functional.
The pairing strengths are fixed separately by fitting the neutron
odd-even mass staggering of the carbon isotopes and they are
$-265$, $-255$, $-325$ $\mathrm{MeV\,fm}^{3} $for 
PC-F1, PC-PK1, and DD-PC1, respectively.
Varying the pairing strengths by about $15\%$ almost does not affect the
conclusions of the present work. Mean-field ground states 
of $^{20,22}$C are pairing-collapsed.

\subsubsection{Projection and GCM}

Each vacuum is projected on angular momentum $J$ using angular momentum projection (AMP) and on neutron and
proton number with particle number projections (PNP)~\cite{Niksic2011_PPNP66-519}. Overlaps of the Bogoliubov
vacua are evaluated with the Pfaffian formula~\cite{Robledo2009_PRC79-021302}. 
The gauge-angle integral of the PNP is a Fomenko sum~\cite{Fomenko1970_JPA3-8} with
$L=7$ neutron gauge angles, for which we find no numerical
instability~\cite{Yao2022_HNP}. 
The axial $K=0$ AMP integral uses $N_{\theta}=12$
Gauss--Legendre nodes.
The AMP+PNP are performed in
the full canonical basis of each DRHBc vacuum, without further
truncation, so that the discretized continuum already present at the
mean-field level is kept in the kernels. This extends the AMP of
deformed continuum states~\cite{Sun2021_SciBull66-1521,Sun2021_PRC104-064319}  to good particle number
and full configuration mixing.

The Hill--Wheeler--Griffin equation is solved in the natural basis of the
norm kernel $N^J$. Natural states with
$n_k/n_{\max}<0.005$ are discarded and we found this truncation can
produce numerical stable results in our calculations.
A generic one-body observable of the yrast GCM state is
$\langle O\rangle^{J}=\sum_{ij}f_i^{J}f_j^{J}O^J_{ij}$, with
$f_i^{J}$ the norm-normalized amplitudes. The collective wave functions
are $g=N^{1/2}f$. Spherical-orbital occupations $n_{nlj}$ are the
Dirac Woods--Saxon spectroscopic occupations, quoted
in the work as filling fractions $n_{nlj}/(2j+1)$. Matter radii are
$R_m=\sqrt{(NR_n^2+ZR_p^2)/A}$. The two-neutron separation energy 
for GCM states is
\begin{equation}
  S_{2n}=E(^{20}\mathrm{C};0^+_1)-E(^{22}\mathrm{C};0^+_1).
\end{equation}

\subsection{Bulk properties}
\label{sec:bulk}

Table~\ref{tab:gcm} lists the PNP+AMP+GCM yrast $0^+$ and $2^+$ of
$^{20,22}$C. Table~\ref{tab:beta} is the generator-coordinate
deformation of those states,
\begin{equation}
  \langle\beta\rangle=\sum_i |g_i|^2\beta_i,\qquad
  \beta_{\mathrm{rms}}=\sqrt{\langle\beta^2\rangle},\qquad
  \sigma_\beta=\sqrt{\langle\beta^2\rangle-\langle\beta\rangle^2}.
\end{equation}
These are the expectation value and fluctuation referred to in the Letter:
oblate deformations dominate the GCM $0^+_1$ of $^{22}$C in all three
functionals. Table~\ref{tab:o24} is the same mix for $^{24}$O.
Table~\ref{tab:mf} gives the pairing-collapsed mean-field minima on the
same decks. Table~\ref{tab:occ} is the $sd$ content of the mixed
$0^+_1$ and $2^+_1$.

\begin{table}[t]
\centering
\caption{PNP+AMP+GCM yrast states.
$E$ in MeV, $B(E2;2^+_1\to 0^+_1)$ in $e^2$~fm$^4$, radii in fm.}
\label{tab:gcm}
\small
\begin{tabular}{llrrrrrrrr}
\toprule
Nuc. & EDF & $E(0^+)$ & $E_x(2^+)$ & $B(E2)$
     & $R_n$ & $R_p$ & $R_m$ & $\langle\beta\rangle_{0^+}$ & $S_{2n}$ \\
\midrule
$^{20}$C & PC-F1
  & $-126.333$ & $2.335$ & $6.11$
  & $3.360$ & $2.607$ & $3.153$ & $-0.385$ & \\
$^{20}$C & PC-PK1
  & $-128.684$ & $3.016$ & $6.23$
  & $3.341$ & $2.535$ & $3.121$ & $-0.254$ & \\
$^{20}$C & DD-PC1
  & $-127.647$ & $3.088$ & $5.09$
  & $3.300$ & $2.626$ & $3.113$ & $-0.285$ & \\
\midrule
$^{22}$C & PC-F1
  & $-127.193$ & $4.123$ & $5.51$
  & $3.515$ & $2.602$ & $3.291$ & $-0.128$ & $+0.86$ \\
$^{22}$C & PC-PK1
  & $-129.248$ & $5.277$ & $4.74$
  & $3.534$ & $2.542$ & $3.293$ & $-0.198$ & $+0.56$ \\
$^{22}$C & DD-PC1
  & $-127.092$ & $5.197$ & $7.58$
  & $3.502$ & $2.620$ & $3.285$ & $-0.162$ & $-0.56$ \\
\bottomrule
\end{tabular}
\end{table}
\begin{table}[t]
\centering
\caption{Collective deformation of the yrast GCM states from $|g_i|^2$.
$\sigma_\beta$ is the fluctuation width.}
\label{tab:beta}
\small
\begin{tabular}{llccccc}
\toprule
Nuc. & EDF
 & $\langle\beta\rangle_{0^+}$ & $\beta_{\mathrm{rms},0^+}$ & $\sigma_{\beta,0^+}$
 & $\langle\beta\rangle_{2^+}$ & $\sigma_{\beta,2^+}$ \\
\midrule
$^{20}$C & PC-F1  & $-0.385$ & $0.563$ & $0.410$ & $-0.512$ & $0.285$ \\
$^{20}$C & PC-PK1 & $-0.254$ & $0.528$ & $0.462$ & $-0.418$ & $0.366$ \\
$^{20}$C & DD-PC1 & $-0.285$ & $0.553$ & $0.474$ & $-0.490$ & $0.324$ \\
$^{22}$C & PC-F1  & $-0.128$ & $0.444$ & $0.425$ & $-0.406$ & $0.343$ \\
$^{22}$C & PC-PK1 & $-0.198$ & $0.445$ & $0.399$ & $-0.388$ & $0.290$ \\
$^{22}$C & DD-PC1 & $-0.162$ & $0.482$ & $0.454$ & $-0.436$ & $0.433$ \\
$^{24}$O & PC-F1  & $+0.041$ & $0.320$ & $0.317$ & $+0.191$ & $0.333$ \\
$^{24}$O & PC-PK1 & $+0.041$ & $0.312$ & $0.309$ & $+0.205$ & $0.320$ \\
$^{24}$O & DD-PC1 & $+0.020$ & $0.305$ & $0.305$ & $+0.237$ & $0.307$ \\
\bottomrule
\end{tabular}
\end{table}
\begin{table}[t]
\centering
\caption{AMP+GCM yrast $0^+$ and $2^+$ of $^{24}$O at
$|\beta|\le 0.8$.
$B(E2)$ in $e^2$~fm$^4$, radii in fm.}
\label{tab:o24}
\small
\begin{tabular}{lrrrrrr}
\toprule
EDF & $E(0^+)$ & $E_x(2^+)$ & $B(E2)$
    & $R_n$ & $R_p$ & $R_m$ \\
\midrule
PC-F1  & $-172.083$ & $6.400$ & $0.003$ & $3.318$ & $2.682$ & $3.121$ \\
PC-PK1 & $-171.963$ & $6.977$ & $0.012$ & $3.335$ & $2.688$ & $3.134$ \\
DD-PC1 & $-172.443$ & $6.919$ & $0.039$ & $3.289$ & $2.687$ & $3.101$ \\
\bottomrule
\end{tabular}
\end{table}
\begin{table}[ht]
\centering
\caption{The total energy $E_{\mathrm{MF}}$, deformations $\beta_{\min}$, neutron matter radius $R_n$, proton matter radius $R_p$, matter radius $R_m$ and separation energies $S_{2n}^{\mathrm{MF}}$, corresponding to the mean-field energy minima in potential energy curve. 
These energies do not depend on $V$.}
\label{tab:mf}
\small
\begin{tabular}{llcccccc}
\toprule
Nuc. & EDF & $\beta_{\min}$ & $E_{\mathrm{MF}}$ (MeV)
     & $R_n$ (fm) & $R_p$ (fm)& $R_m$ (fm) & $S_{2n}^{\mathrm{MF}}$ (MeV)\\
\midrule
$^{20}$C & PC-F1  & $-0.5$ & $-120.499$ & $3.311$ & $2.597$ & $3.114$ & \\
$^{20}$C & PC-PK1 & $-0.4$ & $-121.160$ & $3.285$ & $2.530$ & $3.078$ & \\
$^{20}$C & DD-PC1 & $-0.5$ & $-120.236$ & $3.251$ & $2.623$ & $3.076$ & \\
$^{22}$C & PC-F1  & $\phantom{-}0.0$ & $-122.450$ & $3.391$ & $2.493$ & $3.171$ & $1.95$ \\
$^{22}$C & PC-PK1 & $\phantom{-}0.0$ & $-124.963$ & $3.348$ & $2.395$ & $3.118$ & $3.80$ \\
$^{22}$C & DD-PC1 & $\phantom{-}0.0$ & $-120.874$ & $3.360$ & $2.535$ & $3.156$ & $0.64$ \\
\bottomrule
\end{tabular}
\end{table}
\begin{table}[ht]
\centering
\caption{Woods--Saxon $sd$ occupations of the PNP+GCM yrast states
(particle numbers $n_{nlj}$). Filling fractions $n_{nlj}/(2j+1)$ are
those quoted in the Letter.
$\Delta n=n(^{22}\mathrm{C})-n(^{20}\mathrm{C})$ for the $0^+_1$.}
\label{tab:occ}
\small
\begin{tabular}{llccccccc}
\toprule
Nuc. & EDF & $J^\pi$
     & $n(1d_{5/2})$ & $n(2s_{1/2})$ & $n(1d_{3/2})$ & $n_{sd}$
     & $\Delta n_{d_{5/2}}$ & $\Delta n_{s}$ \\
\midrule
$^{20}$C & PC-F1  & $0^+$ & $4.05$ & $1.43$ & $0.46$ & $5.95$ & & \\
         &        & $2^+$ & $3.95$ & $1.39$ & $0.59$ & $5.94$ & & \\
$^{20}$C & PC-PK1 & $0^+$ & $4.00$ & $1.53$ & $0.36$ & $5.89$ & & \\
         &        & $2^+$ & $3.98$ & $1.47$ & $0.46$ & $5.91$ & & \\
$^{20}$C & DD-PC1 & $0^+$ & $4.23$ & $1.26$ & $0.45$ & $5.95$ & & \\
         &        & $2^+$ & $4.08$ & $1.29$ & $0.58$ & $5.94$ & & \\
\midrule
$^{22}$C & PC-F1  & $0^+$ & $5.36$ & $1.77$ & $0.74$ & $7.86$ & $+1.30$ & $+0.34$ \\
         &        & $2^+$ & $4.79$ & $1.60$ & $1.34$ & $7.72$ & & \\
$^{22}$C & PC-PK1 & $0^+$ & $5.45$ & $1.84$ & $0.41$ & $7.70$ & $+1.45$ & $+0.31$ \\
         &        & $2^+$ & $4.96$ & $1.71$ & $1.00$ & $7.67$ & & \\
$^{22}$C & DD-PC1 & $0^+$ & $5.22$ & $1.68$ & $0.81$ & $7.71$ & $+0.99$ & $+0.42$ \\
         &        & $2^+$ & $4.65$ & $1.51$ & $1.15$ & $7.31$ & & \\
\bottomrule
\end{tabular}
\end{table}

\paragraph{Binding.}
PC-F1 gives a bound PNP
$S_{2n}$ ($+0.86$~MeV) and PC-PK1 is bound by $+0.56$~MeV.
DD-PC1 remains unbound ($-0.56$~MeV).
The mean-field reference states remain bound, so the GCM density of
DD-PC1 is still obtained as a finite-size box state. The same BMF
correlation energy---projection plus mixing lowers $^{20}$C more than
$^{22}$C---reduces $S_{2n}$ in every force.

\paragraph{Radii.}
For the two bound GCM ground states, $R_m(^{22}\mathrm{C})=3.29$~fm
(PC-F1 $3.291$~fm, PC-PK1 $3.293$~fm). DD-PC1 gives a similar numerical
value $3.285$~fm from a finite-size box state.
The carbon-target extraction is $3.44(8)$~fm~\cite{Togano2016_PLB761-412}
[$3.38(10)$~fm in the Glauber reanalysis of
Ref.~\cite{Nagahisa2018_PRC97-054614}]. 
The jump
$\Delta R_m\equiv R_m(^{22}\mathrm{C})-R_m(^{20}\mathrm{C})$ is
$0.14$--$0.17$~fm after MR-DRHBc and only $0.04$--$0.08$~fm at the
mean field, i.e.\ a thick neutron skin rather than a halo at the MF
level. Proton radii barely move ($\Delta R_p\simeq -0.01$ to
$+0.01$~fm); the jump is in $R_n$ ($+0.16$ to $+0.20$~fm). 
Most of
$\Delta R_m$ appears already after angular-momentum projection; GCM
mixing shifts $R_m$ by an additional $\sim 0.03$~fm. From $0^+$ to
$2^+$, $^{20}$C radii are almost unchanged (a rotor). $^{22}$C grows by
$0.06$--$0.10$~fm, consistent with an $sd$ rearrangement.

\paragraph{Spectra and $B(E2)$.}
$^{20}$C is a rotor in all three forces: $E_x(2^+)=2.34$
(PC-F1), $3.02$ (PC-PK1) and $3.09$~MeV (DD-PC1)
(experiment $1.618(11)$~MeV~\cite{Petri2011_PRL107-102501}) and
$B(E2)\simeq 5$--$6~e^2$~fm$^4$ (experiment
$7.5^{+3.0}_{-1.7}$~\cite{Petri2011_PRL107-102501}), with $sd$ occupations that barely
change from $0^+$ to $2^+$. $^{22}$C is not. $E_x(2^+)$ rises to
$4.1$--$5.3$~MeV, the $0^+\to 2^+$ rearrangement is
$n(1d_{5/2},2s_{1/2})\to n(1d_{3/2})$, and $B(E2)$ drops for PC-F1 and
PC-PK1 ($6.11\to 5.51$ and $6.23\to 4.74$) but \emph{rises} for DD-PC1
($5.09\to 7.58$). The PC-PK1 drop is the largest of the three, but the
residual $B(E2)$ is still finite collectivity.
$^{24}$O, by contrast, has $E_x(2^+)=6.4$--$7.0$~MeV and
$B(E2)\simeq 0.003$--$0.039~e^2$~fm$^4$ (Table~\ref{tab:o24}): a
weakly collective $\nu 2s_{1/2}\to\nu 1d_{3/2}$ particle--hole
excitation, the $N=16$ benchmark that $^{22}$C does not reach.

\paragraph{Deformation mixing.}
Table~\ref{tab:beta} makes the same point in $\beta$ space.
$^{20}$C remains oblate ($\langle\beta\rangle_{0^+}=-0.25$ to
$-0.39$) with a wide mix ($\sigma_{\beta,0^+}=0.41$--$0.47$).
$^{22}$C sits closer to the sphere ($-0.13$ to $-0.20$), but
$\sigma_{\beta,0^+}=0.40$--$0.45$ is as large as in $^{20}$C: the
ground state is not pinned at $\beta=0$. PC-F1 is the most spherical
($-0.128$) and does not have the smallest fluctuation.
$^{24}$O is the narrow spherical state
($\langle\beta\rangle_{0^+}\simeq +0.02$ to $+0.04$, $\sigma_\beta\simeq 0.30$).
The $2^+$ of $^{20,22}$C is more oblate and, except for DD-PC1
$^{22}$C, narrower than the $0^+$.

\paragraph{$sd$ configuration.}
The two extra neutrons of $^{22}$C go mainly into $1d_{5/2}$
($\Delta n\simeq +1.0$ to $+1.5$), not into the halo $2s_{1/2}$
($\Delta n\simeq +0.3$). PC-PK1 has the smallest $0^+$ $1d_{3/2}$
occupation ($0.41$, versus $0.74$ and $0.81$) and the $\langle\beta\rangle_{0^+}$
of $-0.198$. That is a relatively spherical, less $d_{3/2}$-contaminated
wave function of the carbon set. It is not an $N=16$ closed shell:
$n(1d_{3/2})$ is finite, $|g(\beta)|^2$ is broad, and the
$2^+$ is an $sd$ rearrangement. In $^{24}$O the same orbitals are
$n(1d_{5/2})\simeq 5.88$--$5.92$, $n(2s_{1/2})\simeq 1.84$--$1.88$,
$n(1d_{3/2})\simeq 0.14$--$0.21$.

\subsection{Density distributions: PC-PK1 and DD-PC1}
\label{sec:supp-den}

\begin{figure}[ht]
\centering
\includegraphics[width=0.8\textwidth]{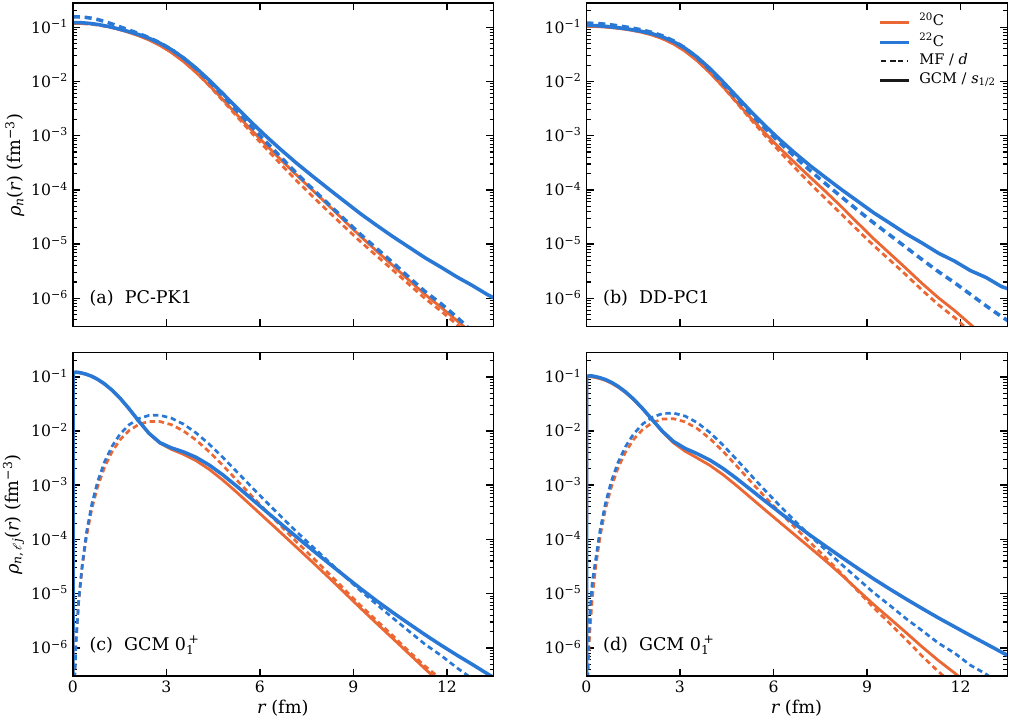}
\caption{Neutron radial densities of $^{20}$C and $^{22}$C
obtained with PC-PK1 [(a),(c)] and DD-PC1 [(b),(d)].
(a),(b)~Spherical monopole density $\rho_n(r)$ for the MF ground state (dashed lines) and for the $0^+_1$ state (solid lines).
(c),(d)~Partial-wave neutron densities $\rho_{n,\ell j}(r)$ in the $0^+_1$ states: the
$s_{1/2}$ component (solid lines) and the $d_{5/2}$ and $d_{3/2}$ components (dashed lines, labeled ``$d$'').}
\label{fig:den0}
\end{figure}

The neutron monopole of the mean-field ground state is compared in
Fig.~\ref{fig:den0} with that of the PNP+GCM $0^+_1$, together with the
GCM $0^+_1$ decomposition into $\ell j$ channels $s_{1/2}$ and
$d=d_{5/2}+d_{3/2}$. PC-F1 is Fig.~1 of the Letter; the two other
functionals are shown here.
The nuclear interior is almost unchanged from $^{20}$C to $^{22}$C. The
extra neutrons appear as a slow $s_{1/2}$ tail beyond $\sim 6$~fm,
consistent with an $s$-wave halo. That tail is a spatial halo in the
$s$ channel of the radial density. 

\subsection{Mean-field and projected PES}
\label{sec:supp-pes}

\begin{figure}[ht]
\centering
\includegraphics[width=0.8\textwidth]{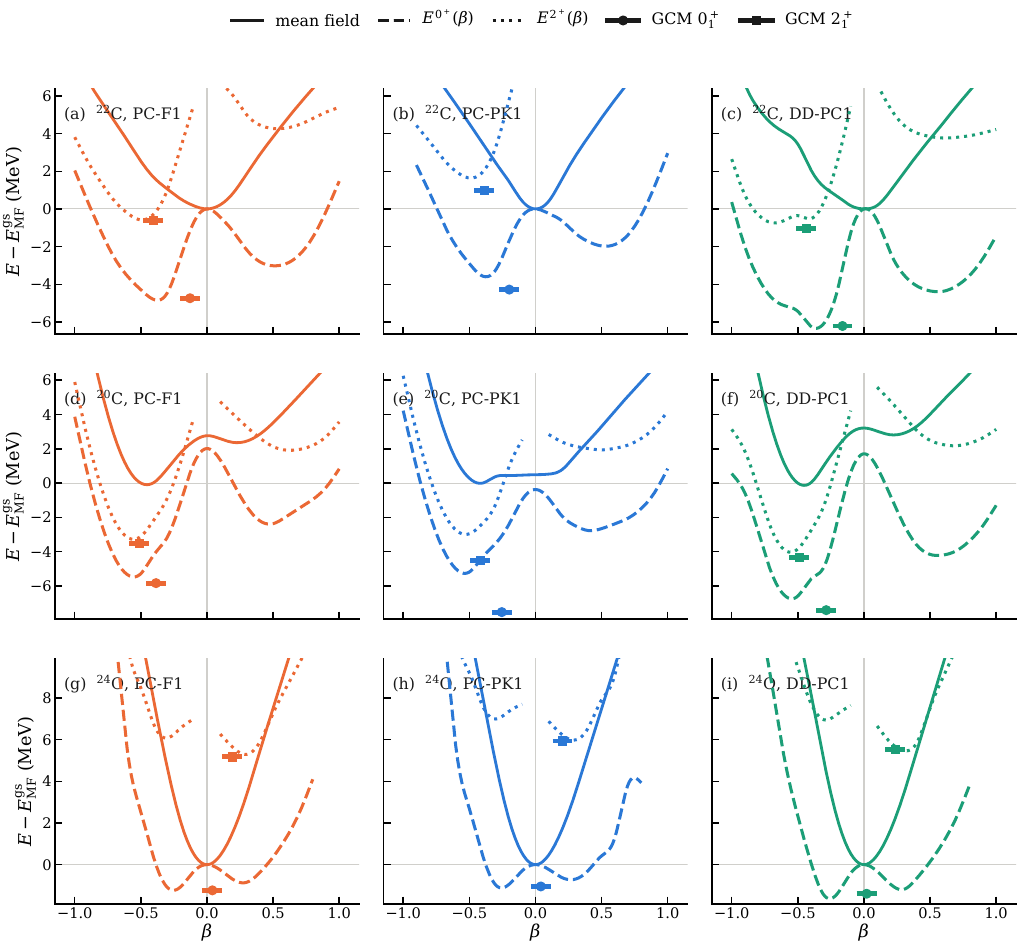}
\caption{Constrained potential-energy surfaces of $^{20}$C,
$^{22}$C, and $^{24}$O obtained with PC-F1, PC-PK1, and DD-PC1.
(a)--(c)~$^{22}$C.
(d)--(f)~$^{20}$C.
(g)--(i)~$^{24}$O.
Mean-field energies (solid lines), diagonal projected $0^+$ energies
(dashed lines), and diagonal projected $2^+$ energies (dotted lines;
the spherical point is omitted) are shown relative to the mean-field
ground state of each panel. Short bars mark the GCM-mixed $0^+_1$
(circles) and $2^+_1$ (squares) states.}
\label{fig:pes}
\end{figure}

Figure~\ref{fig:pes} overlays, for each nucleus and functional, the
mean-field surface $E_{\mathrm{MF}}(\beta)$ and the diagonal projected
energies $E^J(\beta)=\langle H\rangle^J_\beta/N^J_\beta$ ($J=0,2$; no
GCM mixing). All three curves are referred to the mean-field ground-state
energy $E_{\mathrm{MF}}^{\mathrm{gs}}$ of that panel, so the downward
shift of $E^{0^+}$ is the projection correlation energy. 
At the mean field, $^{22}$C and $^{24}$O are spherical in every force
and $^{20}$C is oblate with $\beta=-0.5$. After $J=0$
projection the $^{22}$C minimum moves to $\beta=-0.4$, the same
deformation that dominates the GCM $0^+_1$ weight; the $^{20}$C
minimum stays oblate ($\beta=-0.6$, $-0.5$, $-0.5$); $^{24}$O stays
near spherical with only a
$0.8$--$1.1$~MeV projection gain, versus $3.6$--$6.6$~MeV in the
carbons. The $^{24}$O $2^+$ surface sits $5$--$6$~MeV above the
mean-field ground state; after mixing, $E_x(2^+)=6.4$--$7.0$~MeV
(Table~\ref{tab:o24}). That weakly collective surface is the $N=16$
benchmark. PC-PK1 has the softest $0^+$ surface of
$^{22}$C (lowest spherical barrier in Fig.~\ref{fig:pes}b), which is why
the mixed $|g(\beta)|^2$ is the flattest and $\langle\beta\rangle_{0^+}$
is closest to zero (Table~\ref{tab:beta}). That flatness is shape mixing
of a soft $sd$ wave function, not a rigid spherical $N=16$ closure.

\subsection{Dineutron correlations: PC-PK1 and DD-PC1}
\label{sec:supp-p2}

\begin{figure}[ht]
\centering
\includegraphics[width=\textwidth]{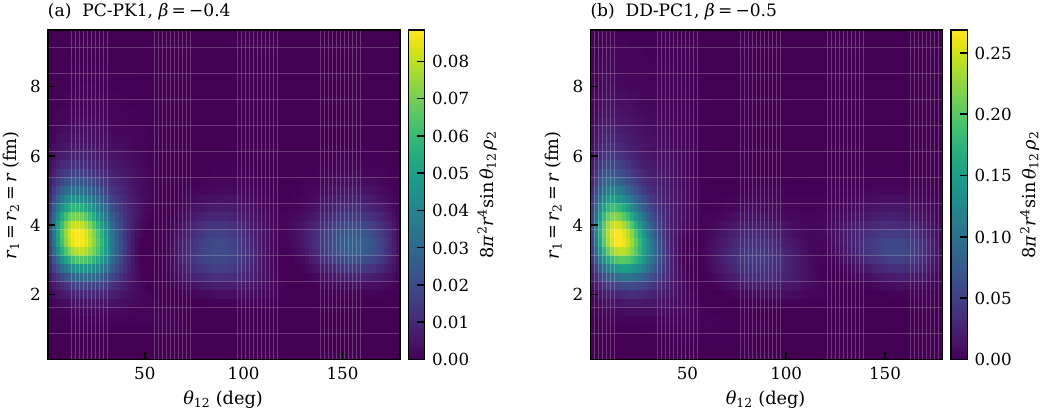}
\caption{Spin-singlet two-neutron correlation density $\rho_2$
of $^{22}$C, evaluated for the $J^{\pi}=0^{+}$ projected generator.
(a)~PC-PK1 at $\beta_{2}=-0.4$.
(b)~DD-PC1 at $\beta_{2}=-0.5$.}
\label{fig:p2}
\end{figure}

The spin-singlet two-neutron correlation density is evaluated for the
$J^\pi=0^+$ projected generator at the oblate peak of $|g(\beta)|^2$,
not on the GCM-mixed $0^+_1$ and not at a mean-field PES minimum.
For PC-F1 that generator is $\beta=-0.4$ (Fig.~4 of the Letter): the
laboratory-frame map peaks at $r=3.5$~fm and $\theta_{12}=18^\circ$.
For PC-PK1 the same generator $\beta=-0.4$ is used. 
The $sd$ occupations
and $|g(\beta)|^2$ (Sec.~\ref{sec:bulk}) are close enough
that the pair-density topology is not expected to change. 
The intrinsic pairing tensor is
\begin{equation}
  \kappa(x_1,x_2)
  =\sum_{\mu>0}u_\mu v_\mu
   \bigl[\psi_\mu(x_1)\bar\psi_\mu(x_2)
        -\bar\psi_\mu(x_1)\psi_\mu(x_2)\bigr],
\end{equation}
and the $S=0$ pair wave function and density are~\cite{Hagino2005_PRC72-44321,Matsuo2005_PRC71-064326}
\begin{equation}
  \Psi_0=\frac{1}{\sqrt{2}}(\kappa_{\uparrow\downarrow}-\kappa_{\downarrow\uparrow}),
  \qquad
  \rho_2=\sum_{\mathrm{spin}}|\Psi_0|^2.
\end{equation}
The physical $J=0$ density is the SO(3) orientation average of the
PNP+AMP kernel, evaluated on the same $(N_\theta,L)=(12,7)$
mesh as the GCM. 
Figure~\ref{fig:p2} shows
$8\pi^2 r^4\sin\theta_{12}\,\rho_2^{J=0}(r,r,\theta_{12})$.
Both functionals are dineutron-dominated. The $J=0$ peak sits at
$r=3.75$~fm and $\theta_{12}=16^\circ$ (PC-PK1) / $14^\circ$ (DD-PC1).
The small-angle and back-to-back weights are
\begin{center}
\small
\begin{tabular}{lcccc}
\toprule
 & \multicolumn{2}{c}{PNAMP $J=0$} & \multicolumn{2}{c}{intrinsic} \\
\cmidrule(lr){2-3}\cmidrule(lr){4-5}
 & $\theta_{12}<60^\circ$ & $\theta_{12}>120^\circ$
 & $\theta_{12}<60^\circ$ & $\theta_{12}>120^\circ$ \\
\midrule
PC-PK1 & $66\%$ & $19\%$ & $69\%$ & $16\%$ \\
DD-PC1 & $74\%$ & $14\%$ & $79\%$ & $10\%$ \\
PC-F1 & $59\%$ & $22\%$ & $65\%$ & $19\%$ \\
\bottomrule
\end{tabular}
\end{center}
As in the work, the dineutron localization arises from the coherent
superposition of opposite-parity $sd$ and $p$
components. Projection reduces the dineutron fraction
relative to the intrinsic state but does not invert the topology.

\end{document}